\documentstyle[aps,pre,epsfig]{revtex}

\begin{document}

\bibliographystyle{unsrt}

\newcommand{\be}{\begin{eqnarray}}
\newcommand{\ee}{\end{eqnarray}}
\newcommand{\xx}{\begin{eqnarray*}}
\newcommand{\yy}{\end{eqnarray*}}
\newcommand{\nn}{\nonumber}
\newcommand{\Vol}{{\rm Vol}}
\newcommand{\sign}{{\rm sign}}
\newcommand{\tr}{{\rm Tr}}
\newcommand{\C}{{\rm Cst\;}}

\title{Magnetic fluctuations in the classical XY model: \\
the origin of an exponential tail in a complex system.}

\vskip 1.0cm

\author{S.T. Bramwell$^1$, J.-Y. Fortin$^{2\ast}$,
P.C.W. Holdsworth$^{3\dagger}$, S. Peysson$^{3}$, J.-F. Pinton$^{3}$,
B. Portelli$^{3}$, and M. Sellitto$^{3}$}

\address{$^1$ Department of Chemistry, University College London,
20 Gordon Street, London, WC1H OAJ, UK
\\
$^2$ Department of Physics, University of Washington, Seattle, USA
\\               
$^3$ Laboratoire de Physique, Ecole Normale Sup\'erieure de Lyon, 46 All\'ee
d'Italie, F-69364 Lyon cedex 07, France
}

\maketitle

%\tighten

\vskip 1cm

\begin{abstract}
We study the probability density function for the fluctuations of the
magnetic order parameter in the low temperature phase of the XY  model 
of finite size.
In two-dimensions this system is critical over the whole of the low 
temperature phase.
It is shown analytically and without recourse to the scaling hypothesis
that, in this case,  the distribution  is non-Gaussian and of universal 
form, independent of both system size and critical exponent $\eta$.
An exact expression for the generating function of the distribution is
obtained, which is transformed and compared with numerical data from 
high resolution molecular dynamics and Monte Carlo simulations.
The asymptotes of the distribution are calculated and found to be of
exponential and double exponential form.
The calculated distribution is fitted to three standard functions:
a generalisation of Gumbel's first asymptote distribution from the 
theory of extremal statistics, a generalised log-normal  distribution, 
and a $\chi^2$ distribution. 
The calculation is extended to general dimension and an exponential tail
is found in all dimensions less than four, despite the fact that critical 
fluctuations are limited to $D=2$.
These results are discussed in the light of similar behaviour observed in
models of interface growth and for dissipative systems driven into a
non-equilibrium steady state.
\end{abstract}

\vskip 2cm


\newpage

%%%%%%%%%%%%%%%%%%%%%%%%%%%%%%%%%%%%%%%%%%%%%%%%%%%%%%%%%%%%%%%%%%%%%%%%%%%%
\section{Introduction}
%%%%%%%%%%%%%%%%%%%%%%%%%%%%%%%%%%%%%%%%%%%%%%%%%%%%%%%%%%%%%%%%%%%%%%%%%%%%

\subsection{Motivation for the Present Work}

The fluctuations in a global measure of a many body system are often
assumed to be of Gaussian form about the mean value~\cite{poi}.
This assumption is nearly always true if the system in question can be
divided into statistically independent microscopic or mesoscopic
elements~\cite{lan.80}, as dictated by the central limit theorem
(see~App.~\ref{app.CLT}).
However, in correlated systems, where this is not the case, there is no
universal reason to expect the central limit theorem to apply.
The fluctuations can then take on a multitude of different mathematical
forms, including those of other, well-defined, limit distributions.

In this context, the most studied correlated systems are critical systems.
At the critical point of a second order phase transition, a correlation 
length, $\xi$, diverges from the microscopic scale (taken as unity through 
out the paper). 
It is only cut off, in an ideal world, by the macroscopic, or integral 
scale $L$.
The probability density function (PDF) for the order parameter $m$ associated 
with the diverging correlation length is essentially the exponential of the 
free energy $P(m) \sim \exp(-F(m)/k_{\rm B}T)$ and takes on an approximately 
Gaussian form as long as the Landau approximation, $F(m) \sim a +b m^2 +...$, 
is valid. Close to the critical point the Landau approximation breaks down 
and the PDF becomes non-Gaussian. 
The key assumption of the renormalisation group theory of 
critical phenomena is that the critical PDF remains scale invariant in 
the thermodynamic limit and can be obtained from the fixed  point of a 
renormalisation group transformation~\cite{cas.78,gar.95} (see 
App.~\ref{app.CLT}).
Thus, renormalisation group theory can be regarded as a generalisation of 
the central limit theorem to systems that are correlated over all length 
scales. 
The critical PDFs can be termed ``universal'', in that, when properly 
normalised,  they depend on at most a few basic symmetries that define 
the  universality class of the system. 
A non-Gaussian and universal PDF is therefore a direct signature of the 
fluctuation  driven critical phenomena that have revolutionised modern 
statistical mechanics~\cite{wil.74}.
Analytical and numerical work~\cite{bru.81,bin.81,binder,bot.00}
on Ising, Potts, and XY models has shown that a generic feature of such 
systems is a skewness, with large fluctuations below the mean, towards 
small order parameter values.

Correlations that are both strong and long range are a feature not only of
critical phenomena but also of systems driven far from equilibrium. 
However, in the case of driven systems, the absence of a microscopic theory 
means that one has to rely heavily on empirical observations from experiment 
and numerical simulation. Labb\'e {\it et al.}~\cite{lab.96} have shown that 
the PDF for the energy injected into a closed turbulent flow at constant 
Reynolds number is also non-Gaussian and universal.
In this case ``universal'' means that the PDF, when suitably normalised,
does not depend on Reynolds number or several other parameters (for example, 
the type of fluid).
The PDF again has a marked skewness with an apparent exponential tail for
fluctuations towards low energies.

The present work is motivated by our empirical observation~\cite{bra.98,pin.00}
that the universal PDF of energy fluctuations in the turbulence
experiment~\cite{lab.96,pin.99} is, within experimental error, of the same
functional form as that of the universal $P(m)$ for the critical system
that we have studied.
The latter is the spin wave limit to the low temperature phase of the 
$2D$-XY model~\cite{arc.97,arc.98} that is known to capture the critical 
behaviour of the full $2D$-XY model~\cite{ber.71,kos.73,kos.74,vil.75,jos.77}.
The distribution is shown in Fig.~\ref{pdf_2d}: it is asymmetric, with 
fluctuations below the mean approaching an exponential asymptote, while 
those above the mean approach a double exponential.
This observation led us to the proposition that many systems, both
equilibrium and non-equilibrium, sharing the property of long range
correlations and multi-scale fluctuations, should share the same features,
at least to a good approximation~\cite{bra.98,bra.00}.
The proposition appears to be strikingly confirmed in ref.~\cite{bra.00}
where, from numerical simulation, similar behaviour is observed in a number
of different systems: for order parameter fluctuations in the two-dimensional
Ising model and in the two-dimensional percolation problem, as well as as for 
fluctuations in global quantities for models of forest fires and avalanches, 
driven into a self-organised critical state.
This appears to contradict the idea that the PDF should depend on the 
particular universality class of the model in hand. 
One possible way of accounting for our observations is that many universality 
classes share common features, with the differences between them appearing 
either outside the range of physical observation, or being hidden by 
experimental error. 
There are therefore many open questions regarding a possible and much 
desired connection between critical phenomena and non-equilibrium
systems as well as regarding the  details of the PDF in critical
systems. 
It is these questions that we address in the current paper, via an analytic 
study of the PDF for  order parameter fluctuations in a finite XY model in 
arbitrary dimension.

\subsection{Normalisation of the Order Parameter}

We discuss order parameter fluctuations of finite systems in terms of
distributions that are calculated in the thermodynamic limit, 
$N \rightarrow \infty$. 
As discussed in App.~\ref{app.CLT}, it is essential to normalise the order 
parameter by an appropriate power of $N = L^D$ in order to obtain a 
distribution  of finite width, or, equivalently, a form for $P(m)$ that is 
independent of system size. 
By extending the scaling hypothesis to include finite systems~\cite{bin.92}, 
the following form of $P(m)$ has been proposed:
\be \label{scale}
P(m,L)\sim L^{\beta/\nu}P_L(mL^{\beta/\nu}, \xi/L).
\ee
Here $\beta$ and $\nu$ are the conventionally defined critical exponents for 
the magnetisation and correlation length $\xi$ respectively~\cite{bin.92}.
The appropriate normalisation of the order parameter is provided by the factor
 $L^{-\beta/\nu}$ while fixing different ratios $\xi/L$ will in principle 
result in an infinity of different limit distributions as 
$L \rightarrow \infty$. 
We concentrate on the case of a truly critical system with correlations over 
all length scales, which should result in maximum deviation from the Gaussian 
form. 
Here the dependence on $\xi$ can be dropped from eqn.~(\ref{scale}), and 
$P_L(m)$ should closely approximate a single universal function of the 
variable $mL^{\beta/\nu}$ for all values of $L$. 
In this form it is independent of the microscopic details of the system, 
although it could indeed depend on the universality class of the transition 
through the critical exponents.

Equation~(\ref{scale}) is demystified somewhat by recognising that the 
normalising factor $L^{-\beta/\nu}$ is, in such an ideal system, 
proportional to the mean value of the order parameter, $\langle m \rangle$.
Further, one of the key properties of a critical system is that the
standard deviation of the distribution, $\sigma$, scales with system size in
the same way as the mean value. 
This property, which is a direct result of the hyperscaling relation and 
which we refer to as the hyperscaling condition, means that 
eqn.~(\ref{scale}) can alternatively be written in the form 
$P(m) = 1/\sigma P_L(m/\sigma)$.
Thus $\sigma$ provides, as might be intuitively expected, the correct 
normalisation of the order parameter, such that a reasonable PDF of finite 
width is obtained in the limit $N \rightarrow \infty$.
In this paper, in addition, we shift the distribution with respect to the
mean value and define
\be \label{scale-bhp}
\sigma P(m) & = & \Pi (\theta) \,,
\ee
where
\be
\theta =  \frac{m-\langle m \rangle}{\sigma}.
\ee
In this representation one expects the PDF to fall, in the thermodynamic
limit, onto a single universal curve.
Provided that finite-size corrections to scaling are negligible one should 
observe data collapse onto this universal curve for large but finite system 
sizes.

\subsection{The Two Dimensional XY Model}

The model that we study, the harmonic spin wave limit to the XY model, is
defined in section~\ref{sec.pdf_2}
for the case of two dimensions, $D = 2$.
This is the dimension of most interest in the present context, as the system 
is at its lower critical dimension. 
At low temperature the coupling $J/k_{\rm B}T$ 
is an exactly marginal variable that characterises a line of critical points 
in zero applied field~\cite{car.96}.
The critical line is separated from the paramagnetic phase by the
Kosterlitz-Thouless-Berezinskii phase transition at 
$T_{\rm KTB}$~\cite{ber.71,kos.73}.
The critical phase that exists below this temperature is an attractive subject
of investigation from both an analytic and a numerical point of view.
Its physics is entirely captured by the harmonic
Hamiltonian~\cite{ber.71,vil.75,jos.77} with the result that many
calculations can be performed microscopically, without the need to use 
renormalisation techniques, or the scaling hypothesis.
From a numerical point of view simulation results near a single, isolated,
critical point are often complicated by a shift in the effective critical
temperature by an amount scaling to zero as 
$L^{-1/\nu}$~\cite{bin.81,binder,bot.00}, making it unclear exactly which 
temperature should be studied.
Indeed numerical studies of Ising and Potts models~\cite{bin.81,binder,bot.00} 
do find distributions whose form depends on temperature in the critical 
region~(see App.~\ref{app.CLT}).
In the $2D$-XY model, as the system is critical over a range of temperatures 
there are no such technical problems and data for $P(m)$ can be collected at 
all points below $T_{\rm KTB}$.
These factors make the $2D$-XY model an ideal system with which to study
the effects of critical correlations.

The finite-size scaling for the $2D$-XY model has been discussed in our
previous publications~\cite{arc.97,arc.98}.
In this work, we began, following Berezinskii~\cite{ber.71} and R\'acz and
Plischke~\cite{rac.94}, an exact calculation for the PDF of order parameter
fluctuations.
This calculation is completed and presented in detail in the current paper
(section~\ref{sec.pdf_2}).
It shows explicitly that the non-Gaussian behaviour in the $2D$-XY model stems
from the influence of all length scales from the microscopic to the macroscopic
scale.
We propose that the same is true for other complex systems including those
driven far from equilibrium, which provides a basis for understanding the 
apparent overlap of their PDFs and provides an unexpected experimental 
motivation for studying a system as simple as the $2D$-XY model.

Two results coming out of our calculation are worthy of note at this stage.
The first is an exact analytic result that is rather surprising given the 
previous discussion and the general belief concerning the dependence of the 
PDF on universality class: shifting the curve with respect to the mean,
eqn.~(\ref{scale-bhp}), gives us universal data collapse, not only for all
system size but also for all temperatures for which the harmonic Hamiltonian 
is valid.
The ratio of exponents $\beta/\nu$ depends linearly on temperature, from which
we deduce that the PDF is independent of the value of the exponents along
the line of critical points.
One should note however that these points are rather special and the result
cannot necessarily be generalised to all critical points: not all the 
usual critical exponents are defined.
For example the exponents $\beta$ and $\nu$ are not individually defined,
but their ratio is~\cite{kos.73}.
The usual scaling relations are valid in terms of the ratio $\beta/\nu$ only  
and this ``weak scaling''~\cite{kos.74} means that there is only one
independent critical exponent, $\eta = 2\beta/\nu$~\cite{kos.73}, compared 
with two for a regular critical point.
This is all that is required for the analysis leading to 
eqn.~(\ref{scale}), but is not sufficient to ensure a unique functional 
form for the general problem with two exponents. However,
it does seem consistent with the idea that only small differences
separate results for different universality classes.

The second result that it is relevant to mention at this stage concerns the
finite size scaling data collapse of eqn.~(\ref{scale-bhp}). We find that the
hyperscaling property $\sigma
\sim \langle m \rangle$ is not a necessary condition for data collapse onto
a non-Gaussian function.
With our definition (\ref{scale-bhp}), the first two moments of $P(m)$ fall
trivially out of the calculation and all that is required for
data collapse is that the moments $\langle \theta^p \rangle$ for $p\; >\; 2$, 
are independent of the system size. 
This is the most general condition for non-Gaussian data collapse, while the 
PDF only satisfies the scaling hypothesis in the form of eqn.~(\ref{scale}),
if the hyperscaling condition is satisfied. 
We give, in section II.A, an explicit example where data collapse onto the 
universal curve of the $2D$-XY model occurs, but where the hyperscaling 
condition is not satisfied. 
If we make an expansion of the order parameter about a perfectly
ordered state ($m=1$) in powers of
temperature, keeping only the linear term, then $\langle m \rangle$ diverges
logarithmically with system size~\cite{mer.68,tob.79},
while the standard deviation is a constant.
The ratio $\sigma/\langle m \rangle$ is actually an {\it increasing} function of
system size throughout the physical domain. It is only when
the order parameter is correctly defined on the interval $[0,1]$ that
the hyperscaling relation is re-established, but written in the form
(\ref{scale-bhp}) the two distinct variables have the same universal PDF,
even outside the range of temperature and system size for which the
development gives an accurate representation of the true order parameter.
This result is more than a mathematical curiosity; the harmonic approximation
for the $2D$-XY model maps directly onto the Edwards-Wilkinson (EW)
model~\cite{edw.82,fol.94,pli.94,Antal,pim.99} for interface growth and
the linearised order parameter is related to the square of the
interface width, $w$: $m = 1- w^2$. 
Our PDF therefore corresponds precisely to that for interface width 
fluctuations and for which, in two dimensions, the hyperscaling condition 
for the observable $w^2$ is explicitly violated.

\subsection{Organisation of the Paper}

The rest of the paper is organised as follows. 
In section~\ref{sec.pdf_2} we present detail of the calculation for the 
PDF in the $2D$-XY model~\footnote{
For convenience, throughout this paper we use the term ``XY model'' to
refer to either the model defined by the spin wave Hamiltonian or the full
XY model over a temperature range in which the spin wave approximation is
valid. 
This should not cause any confusion in reading the present paper, but our 
choice of terminology should be born in mind when comparing to other work 
on the XY model}.
We show explicitly that it is a universal function of system size and of
temperature and find an exact expression for the characteristic function
(section~\ref{subsec.pdf_analytic}).
Transforming the distribution numerically, we compare it in detail with
extensive Monte Carlo and molecular dynamics simulations of the full XY
model and show that it is clearly the complete solution of the problem
(section~\ref{subsec.pdf_simulation}). 
We calculate the asymptotic values of the distribution for large deviations
below and above the mean, which we find to be exponential and double
exponential respectively
(sections~\ref{subsec.pdf_lop}, and~\ref{subsec.pdf_asymptotes}).

In section~\ref{sec.fitting} we try to fit the computed PDF to standard 
functions by comparing the moment expansion of the generating function 
with those of the Fourier transform of the test function. 
Three functions are considered:
\be\label{fit}
 \Pi(\theta) &\sim & \left\{
\begin{array}{ll}
\exp a\left[\theta-s -\exp(\theta-s)\right] \,,  \\ \\
\frac{1}{s-\theta}
\exp \left\{-\left[\log(s-\theta)-a \right]^2 \right\}  \,,   \\ \\
(s-\theta)^{\nu/2-1}{\rm e}^{-a(s-\theta)}   \,.
\end{array}
\right.
\ee
The PDF is fitted to an excellent approximation over the physical range by 
the first two functions, while the third give a reasonable but slightly 
inferior fit.
The first function, with $a$ an integer, comes from extremal statistics
(section~\ref{subsec.gumbel}). 
It is Gumbel's third asymptote, corresponding to the PDF for the
$a^{th}$ largest value from ensembles of $N$ random numbers~\cite{gum.58}.
The interpretation with $a$ non-integer (we find $a=\pi/2$) is not clear, 
but a connection between critical phenomena and extremal statistics is a 
very appealing concept~\cite{bou.97,cha.00}. 
The second function is a generalised log-normal distribution
(section~\ref{subsec.lognormal}).
Unlike the first curve, it does not have the correct asymptotic forms but 
despite this it fits just as well over the physical domain.
The third function is a $\chi^2$ distribution describing identical and
statistically independent degrees of freedom (section~\ref{subsec.chisquare}). 
It gives reasonable qualitative agreement indicating that a good, zeroth 
order description of a correlated system is in terms of a reduced number 
of statistically independent variables. 
However, this description has its limits, as shown by the fact that this 
function fits the exact PDF slightly less well than the other two.
This variety of different fits suggests that one should treat the physical 
interpretations that they offer with caution; however even with this caveat 
in mind they still represent useful mathematical tools.
To investigate this point further, in section~\ref{subsec.pearson} we 
derive an approximate  functional form for the curve using an analysis 
due to Pearson which reconstructs the PDF from the four principle moments, 
which in this case have been calculated analytically. 
The Pearson analysis gives a quite different function
which also gives a good description of the exact PDF over a physical
range  of $\theta$. 
This serves to emphasise that, given zero mean and unit variance, the shape 
of the curve over a typical experimental range is essentially  defined by 
its skewness, $\gamma$, and kurtosis, $\kappa$. 
Therefore, an alternative way of summarising the observed 
universality~\cite{bra.98,bra.00}, is that $\gamma$ and $\kappa$, 
for several different systems, have the same scale-invariant values as 
they do for the XY model.

In section~\ref{sec.pdf_D} we extend our calculation to $D$ dimensions, 
which apart from $D = 2$ are all non-critical. 
Despite this, we find evidence of the integral scale for all dimensions 
$D<4$.
For $D=1$, the PDF for the linearised order parameter shows an exponential 
tail. However, we show numerically that the PDF for the correctly defined 
order parameter is quite different and is just what one would expect for a 
paramagnetic system without correlations 
(section~\ref{subsec.pdf_1}).
The case $D=3$ holds a final surprise (section~\ref{subsec.pdf_3}): 
despite the long range order of the low temperature phase the PDF
is still not a Gaussian function.
The temperature is a dangerously irrelevant variable in the ordered 
phase of the $3D$-XY model with the result that the susceptibility remains 
weakly divergent at low temperature~\cite{hol.40}. 
The result of this divergence is that the asymptotes of the PDF for large 
fluctuations are exponential below the mean and $\exp(-\C \theta^3)$ above
the mean. 
The hyperscaling relation, in this case, is again violated.
The divergence disappears at the upper critical dimension and we find a 
truly Gaussian PDF for $D \ge 4$.

In section~\ref{sec.conclusion} we conclude by returning to the physical 
reasons for the exponential tail in the PDF. 
The XY model is diagonisable in reciprocal space reducing it to a model 
of statistically independent degrees of freedom: spin wave 
amplitudes at wave vector ${\bf q}$, $\phi_{\bf q}$.
The amplitudes  $\langle \phi^2_{\bf q}\rangle $ diverge at small $q$
and are these modes that give the non-Gaussian fluctuations.
In one-dimension they completely destroy magnetic order, in two-dimensions
they give critical behaviour and between two and four dimensions they give
remnant critical behaviour in the form of a dangerously irrelevant variable.

%%%%%%%%%%%%%%%%%%%%%%%%%%%%%%%%%%%%%%%%%%%%%%%%%%%%%%%%%%%%%%%%%%%%%%%%%%%%%%
\section{Probability density function for the order parameter in
         the $2D$-XY model}
\label{sec.pdf_2}
%%%%%%%%%%%%%%%%%%%%%%%%%%%%%%%%%%%%%%%%%%%%%%%%%%%%%%%%%%%%%%%%%%%%%%%%%%%%%%

%%%%%%%%%%%%%%%%%%%%%%%%%%%%%%%%
\subsection{Analytic Expression}
\label{subsec.pdf_analytic}
%%%%%%%%%%%%%%%%%%%%%%%%%%%%%%%%

The $2D$-XY model is defined by the Hamiltonian
\be \label{eq1}
H=-J\sum_{\langle i,j \rangle}\cos(\theta_i-\theta_j),
\ee
where the angles $\theta_i$
refer to the orientation of classical spins ${\bf S}_i$ confined in a
plane and where the sum is over nearest neighbour spins.
In the following we consider a square lattice of side $L$, with
periodic boundaries.  The order parameter is a two dimensional
vector ${\bf m}$ which, in zero field is free to point in any
direction. We define the instantaneous magnetisation  as
the scalar $m = |{\bf m}|$
\be \label{eq-m}
m = {1\over{N}} \sum_{i=1}^N \cos(\theta_i - \overline{\theta}),
\ee
where
$\overline{\theta} = \tan^{-1}
(\sum_i \sin \theta_i/\sum_i \cos \theta_i)$ is
the instantaneous magnetisation direction. Within small corrections,
which disappear in the thermodynamic limit,
this corresponds to the more conventional definition
$$
m=\frac{1}{N}\sqrt{\left ( \sum_{i=1}^N{\bf S}_i \right )^2.}
$$
For all temperatures below  $T_{\rm KTB}$  the renormalisation group 
trajectories flow, at large length scale, towards a regime where only 
spin-wave excitations are relevant~\cite{vil.75,jos.77}. 
The physics of the low temperature phase is therefore completely captured 
by the quadratic Hamiltonian
\be \label{eq2}
H={J\over{2}}\sum_{\langle i,j \rangle}\left (\theta_i-\theta_j\right )^2.
\ee
We therefore restrict ourselves, in the following calculation to this
Hamiltonian and neglect the periodicity of the variables $\theta_i$. 
Our calculation can not therefore take into account the presence of 
vortex pairs. 
Close to $T_{\rm KTB}$ in two dimensions and also in one dimension, 
where free vortices are relevant variables we would expect a deviation 
from the behaviour shown in Fig.~\ref{pdf_2d}.
This point is discussed further below.

We now calculate the PDF $P(m)$ that the system be in a state with 
magnetisation $m$, using the standard property that a probability 
density function is entirely defined by the value of its 
moments~\cite{fel.71}. 
Indeed, $P(m)$ can be expressed in terms of its characteristic function,
$\tilde{P}(x)$:
\be
P(m) = \int_{-\infty}^{\infty} \frac{dx}{2\pi}\;{\rm e}^{imx} 
\tilde{P}(x) \,,
\ee
which can in turn be expanded in a Taylor series whose coefficients
are the moments $\langle m^p \rangle $ :
\be\label{moments}
\tilde{P}(x) = \sum_{p=0}^{\infty} \frac{x^p}{p!}
\left. \frac{\partial^{p} \tilde{P}}{\partial x^p} \right|_{x=0} =
\sum_{p=0}^{\infty} \frac{(-ix)^p}{p!} \langle m^p \rangle \; ,
\ee
so that
\be\label{eq4}
P(m)=\int_{-\infty}^{\infty} \frac{dx}{2\pi}\;
{\rm e}^{imx}\sum_{p=0}^{\infty}\frac{(-ix)^p}{p!} \langle m^p \rangle \,.
\ee
Eqn.~(\ref{eq4}) assumes that the series converges and that all the moments 
exists. 
Note that this last feature demands that $P(m)$ falls off 
faster than any power-law of $m$.

The program for calculating $P(m)$ is therefore to calculate the moments
$ \langle m^p \rangle$, sum the series and transform the final result.
To this end it is useful to define the Green function in Fourier space,
\be \label{eq3}
G({\bf q}) & = & \frac{1}{4-2\cos q_x-2\cos q_y} \,,
\ee
where $q_x$ and $q_y$ take the discrete values $\frac{2\pi}{L} n$ of the
Brillouin zone with $n=0,\ldots L-1$. We also define the set of constants
 $g_k=\sum_{{\bf q}}G({\bf q})^k/N^k$. The value of $g_1$ diverges
logarithmically with system size, illustrating
the critical nature of the low temperature phase~\cite{jos.77,tob.79}:
$g_1 = {1\over{4\pi}} \log(CN), \; C = 1.8456$~\cite{C}.
The values of $g_k,\;k\ge 2$ are independent of $N$ in the thermodynamic
limit.
We find:
$g_2\simeq 3.8667\;10^{-3}$,
$g_3\simeq 7.5719\;10^{-5}$,
$g_4=1.7626\;10^{-6}$
and that for large $k$, $g_k$ behaves like $(2\pi)^{1-2k}/2(k-1)$, 
see App.~\ref{app.1}.

The first moment is easily calculated within this approximation (see
App.~\ref{app.1} and refs.~\cite{tob.79,arc.97}).
One finds that $\langle m \rangle $ decreases algebraically with the 
size, as one would expect from finite-size scaling~\cite{bin.92}
\be  \label{eq2b}
 \langle m \rangle & = & (NC)^{-k_{\rm B}T/8\pi J} \;\;.
\ee
As discussed above, while the critical exponents $\beta$ and
$\nu$ are not individually defined for the $2D$-XY model, their ratio
is~\cite{kos.74} and the system obeys what Kosterlitz referres to as
weak scaling~\cite{kos.74}. Through eqn.~(\ref{eq2b}) the
ratio of exponents is defined: $\beta/\nu = \eta/2 = T/4\pi J$.

\begin{figure}
\begin{center}
\epsfig{file=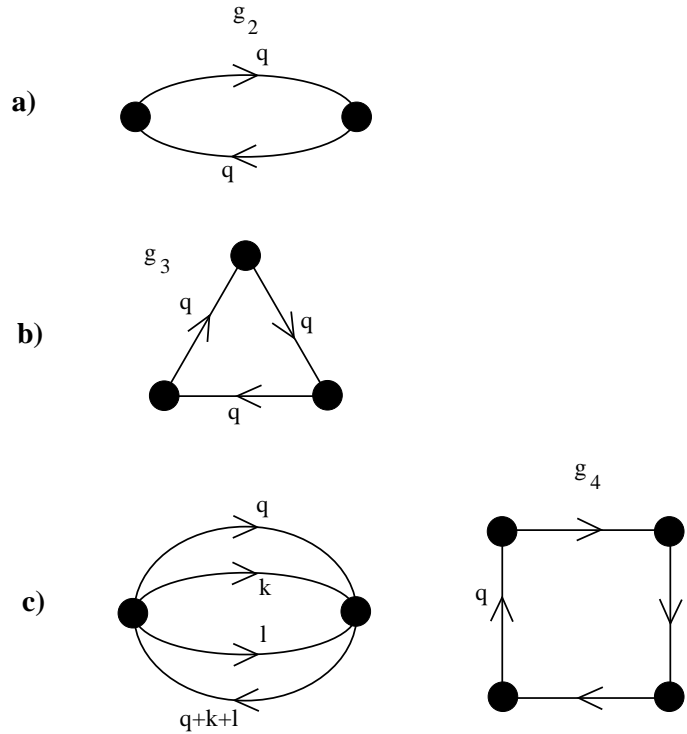,width=9cm}
\end{center}
\bigskip
\caption{Diagrams contributing to the distribution: 
  a) to order $\tau^2$, 
  b) to order $\tau^3$ and 
  c) to order $\tau^4$.}
\label{diagrams}
\end{figure}

For higher moments we need a more systematic approach.
A specific property of the quadratic Hamiltonian (\ref{eq2}) is
that the moments can be calculated using the tools of Gaussian
integration~\cite{arc.97,gawedski}.
In particular, by the application of Wick's theorem, propagators of order
$2p$ in reciprocal space can be exactly expressed in terms of quadratic
propagators so that the $p^{th}$ moment is proportional to
${\langle m \rangle}^p  $.
One finds~\cite{arc.98,pey.97}
\be\label{eq5}
\langle m^p \rangle
& = &
{\langle m \rangle}^p \frac{1}{(2N)^p}  \sum_{{\bf r}_1,\ldots,{\bf r}_p}
\sum_{\sigma_1,\ldots,\sigma_p = \pm 1}\exp\left[
-\frac{\tau}{2}\sum_{i\neq j}\sigma_i G_R({\bf r}_i-{\bf r}_j)
\sigma_j \right]  \,,
\ee
where $\tau$ is the reduced temperature
$k_{\rm B}T/J$ and $G_R({\bf r})$
the regularised Green function
$\sum_{q\neq 0}G({\bf q})\exp(i{\bf q} \cdot {\bf r})/N$.
In order to compute each moment of order $p$, we have to evaluate the
sums over the positions and operators $\sigma_i$. The idea is to expand
the exponential term~(\ref{eq5}) and introduce a diagrammatic representation
of each quantity computed. For example, we represent $\sigma_i
G_R({\bf r}_i-{\bf r}_j)\sigma_j$ by a line between $i$ and $j$ on
a lattice of $p$ sites. The general term of the expansion is then a
set of graphs with a combinatorial factor for the symmetries.
Since $\sigma_i^2=1$, only closed diagrams are relevant, the factor $2^p$ being
cancelled by the sum over all the $\sigma_i$. The factor of $\tau^k$, is
common to all graphs with $k$ lines connected together, with an even
connectivity at each vertex.
For example, up to the second order term in $\tau$, we have
\be
\langle m^p \rangle
& = &
{\langle m \rangle}^p\left[
1+\left(\frac{-\tau}{2}\right )^2\frac{1}{2!}2p(p-1)
\frac{1}{N}\sum_{{\bf r}}G_R^2({\bf r}) + \cdots
\right] .
\ee
The term $\sum_{{\bf r}}G_R^2({\bf r})/N^2=\sum_{{\bf q}\neq
{\bf 0}}G({\bf q})^2/N^2=g_2$ is the value of the one loop graph with
two lines, as shown in  Fig.~\ref{diagrams}a. 
There is an additional  symmetry factor $2\times p(p-1)$, which is the 
number of possible positions for such diagrams connecting 2 lines on a 
closed graph on a lattice of $p$ points. 
For the third order term in $\tau$, we have only one diagram with 3 
vertices, of value  $g_3$, Fig.~\ref{diagrams}b.  The symmetry factor is 
equal to $p(p-1)(p-2)\times 4\times2$. 
The factor $4\times 2$ comes from the number of possible ways of connecting 
3 lines together. 
For the $4^{th}$ order term there are three different graphs, two of which 
are shown in Fig.~\ref{diagrams}c. 
The first has 3 loops and 2 vertices, the second, of value $g_4$, has one 
loop and 4 vertices.
The third graph, not shown, consists of two disconnected one loop graphs
of the type shown in Fig.~\ref{diagrams}a.
In general, at each order in $\tau$, we  have the product of different 
closed diagrams, with one or many loops. 
It appears that the values of multiple loop graphs, like the first one in 
Fig.~\ref{diagrams}c), are zero in the thermodynamic limit. 
We therefore find that only the one loop diagrams are relevant and the value 
for such a diagram, with $k$ lines and $k$ vertices is $g_k$.
We can now express the $p^{th}$ moment of the magnetisation as
\be\label{eq6}
\frac{  \langle m^p \rangle }{ \langle m \rangle ^p}
& = & 1+\sum_{k\geq 2}\left (\frac{-\tau}{2}\right )^k
\frac{1}{k!}\sum_{r\geq 1}\sum_{
\stackrel{k_1+\cdots+k_r=k}{k_i\geq 2}}
g_{k_1}\cdots g_{k_r}C(k_1,\cdots,k_r)\times p(p-1)\cdots (p-k+1) \,,
\ee
with $C(k_1,\ldots,k_r)$ a combinatorial factor which takes into account
the possible ways of putting together $k$ lines on $r$ graphs,
the first with $k_1$ lines, the second with $k_2$ lines, etc...,
including the symmetries. For example, the factor associated with
one triangle is $C(3)=4\times 2$. It is then relatively easy to show that
\be\label{eq7}
C(k_1,\ldots,k_r)=\frac{2^{k-r}k!}{(k_1+\cdots+k_r)(
k_2+\cdots+k_r)\cdots k_r}  \,.
\ee
Next, we can use the fact that every diagram is invariant by the action of
the group ${\cal S}_r$ of permutations of its $r$ single elements, so that,
instead of~(\ref{eq7}), one can use a more convenient form for the
combinatorial factor:
\be
\frac{1}{r!}\sum_{\sigma\in {\cal S}_r}C(k_{\sigma(1)},\ldots,k_{\sigma(r)})
=\frac{1}{r!}\frac{2^{k-r}k!}{k_1\cdots k_r}  \,.
\ee
Setting $f_{k_i}=g_{k_i}(-\tau)^{k_i}/2k_i$ we arrive at the result
\be\label{exp}\nn
\frac{\langle m^p\rangle}{{\langle m \rangle}^p }
& = &
1+\sum_{k\geq 2}\sum_{r=2}^k
\frac{1}{r!}
%\sum_{\stackrel{k_1+\cdots+k_r=k}{k_i\geq 1}}
\sum_{k_1+\cdots+k_r=k}
f_{k_1}\cdots f_{k_r}\times p(p-1)\cdots (p-k+1)
\\ \nn
&=&1+\sum_{k\geq 2}\sum_{r=2}^k
\frac{1}{r!}
%\sum_{\stackrel{k_1+\cdots+k_r=k}{k_i\geq 1}}
\sum_{k_1+\cdots+k_r=k}
f_{k_1}\cdots f_{k_r}\left .\frac{\partial^k}{\partial z^k}
z^p \right |_{z=1}
\\
&=& \left.\exp \left[\sum_{k=2}^{\infty}\frac{g_k}{2k}(-\tau)^k
\partial_z^k \right] z^p \right|_{z=1}.
\ee
For $p=2$ we find $\langle m^2 \rangle /\langle m \rangle ^2 =1+g_2
\tau^2/2$ and defining
$\sigma=\sqrt{\langle m^2 \rangle - \langle m \rangle^2}$  we thus
arrive at the hyperscaling condition that the ratio  $\sigma/\langle m\rangle$
is independent of the system size. Hence
\be\label{sig}
\sigma & = &
\sqrt{\frac{g_2}{2}}\frac{k_{\rm B}T}{J}  \langle m \rangle    \,.
\ee
One can now substitute for $ \langle m^p \rangle $ in~(\ref{eq4}) 
using~(\ref{exp}) and after re-arranging the summations the distribution 
can finally be expressed as an integral, depending on the values of the 
one loop diagrams $g_k$ only
\be\nn
P(m)=\int_{-\infty}^{\infty}\frac{dx}{2\pi}
\exp \left[ ix \left( m-  \langle m \rangle \right)
+\sum_{k=2}^{\infty}\frac{g_k}{2k}
(i\tau   \langle m \rangle x)^k \right] \,.
\ee
Changing variables,  $x \rightarrow x/\sigma$ and using (\ref{sig}) we find
\be\label{eq9}
P(m)=\int_{-\infty}^{\infty}\frac{dx}{2\pi\sigma}
\exp\left [ix \frac{m- \langle m \rangle}{\sigma}
+\sum_{k=2}^{\infty}\frac{g_k}{2k}
\left (ix\sqrt{\frac{2}{g_2}}\right )^k \right ],
\ee
which is the principal result of ref.~\cite{arc.98}.
Defining $\sigma P(m)=\Pi(\theta)$, we see that the function $\Pi$
depends uniquely on the variable $\theta = (m-\langle m \rangle )/\sigma$
and the $g_k,\; k\ge2$.
As the $g_k$ are constants in the thermodynamic limit, $\Pi(\theta)$
is a universal function, independent of both system size and temperature.  
The asymmetry comes from the fact that the ratios  
$g_k/(g_2/2)^{k/2}, \; k\ge 3$  are non zero and this constitutes the 
abnormal influence of the integral scale. 
If, in the thermodynamic limit, $k=2$ were the only non-zero term one would 
arrive at a Gaussian PDF centered on $\langle m \rangle $. 
Departure from a Gaussian function is typically characterised by the skewness,
 $\gamma = \langle \theta^3 \rangle$ and kurtosis, 
$\kappa = \langle \theta^4 \rangle $~\cite{bur.99}. 
We find
\be\label{skewness}
\gamma &=& - {g_3\over{(g_2/2)^{3/2}}} = -0.8907 \,, \nn \\
\kappa &=& 3 + 3{g_4\over{(g_2/2)^2}} = 4.415 \,.
\ee
%
%%%%%%%%%%%%%%%%%%%%%%%%FIGURE%%%%%%%%%%%%%%%%%%%%%%%%%
%%%%%%%%%%%%%%%%%%%%%%%%%%%%%%%%%%%%%%%%%%%%%%%%%%%%%%%%%%%%%%%%%
%
Although we can calculate the asymptotic behaviour of $g_k$ for large
$k$, we are not able to compute the constants analytically and so we 
cannot sum the series (\ref{eq9}). 
However we can transform it into a very much more useful form by keeping 
$N$ large but finite and inverting the sums over ${\bf q}$ and $k$. 
The even and odd terms are separated and summed independently and we 
eventually find:
\be\label{eq10}
\Pi(\theta)=\int_{-\infty}^{\infty}\sqrt{\frac{g_2}{2}}\frac{dx}{2\pi}
\exp\left\{
i x \theta\sqrt{\frac{g_2}{2}} - 
\sum_{{\bf q}\neq {\bf 0}}
\left[ \frac{i}{2} xG({\bf q})/N 
- \frac{i}{2} \arctan \left( xG({\bf q})/N \right) 
+\frac{1}{4} \log \left(1 + x^2 G ( {\bf q})^2/N^2 \right) 
\right] \right\} .
\ee
The sum over ${\bf q}$ and the integral over $x$, in~(\ref{eq10}) 
can now be performed numerically, allowing the evaluation of 
$\Pi(\theta)$.

%%%%%%%%%%%%%%%%%%%%%%%%%%%%%%%%%%%%%%%
\subsection{Comparison with Simulation}
\label{subsec.pdf_simulation}
%%%%%%%%%%%%%%%%%%%%%%%%%%%%%%%%%%%%%%%

To test the above calculation and to verify its scaling properties we 
have carried out extensive numerical simulations of the $2D$-XY model
with full cosine interaction, eqn.~(\ref{eq1}), for different values of 
temperature and system size. 
In addition, we have also done microcanonical molecular dynamics (MD) 
simulations to check the possible dependence of the PDF for fluctuations 
on the statistical ensemble.

The Monte Carlo simulations were performed with $10^8$ Monte Carlo steps
per spin, with $10^6$ steps used for equilibration.
The MD simulation was carried out for systems of $N$ classical 
rotators,~\cite{leo.98}, with Hamiltonian
\be
H_R = \sum_{i=1}^N {\dot\theta_i^2\over{2}}  +
               J \sum_{\langle ij \rangle}
               \left[ 1- \cos( \theta_i - \theta_j) \right] \,.
\ee
The equations of motion were integrated numerically, using a Verlet algorithm.
In order to explore the low-temperature fluctuations regime the initial
configuration of the system was chosen with the spins pointing in the same
direction and with a Gaussian distribution of momenta.
The system was then equilibrated for a time of $10^6-10^7$ sweeps
and data collected over a time span of $10^8-10^9$ sweeps according
the size of the system.
Note that one cannot use the harmonic interaction~(\ref{eq2}) to study
deterministic dynamics in the microcanonical ensemble, as this would allow
no coupling between the spin  wave modes and no evolution would be possible.
The non-linearity of the cosine interaction allows mixing between the
normal modes and the sampling of equilibrium states.
Here we do not report work at high enough energies to allow vortex
formation~\cite{jen.91,sel.00} with any significant probability.
Rather, the non-linearity plays the role of the heat bath in the canonical
ensemble, while the physics is still correctly described by the harmonic part
of the interaction.

\begin{figure}
\begin{center}
\epsfig{file=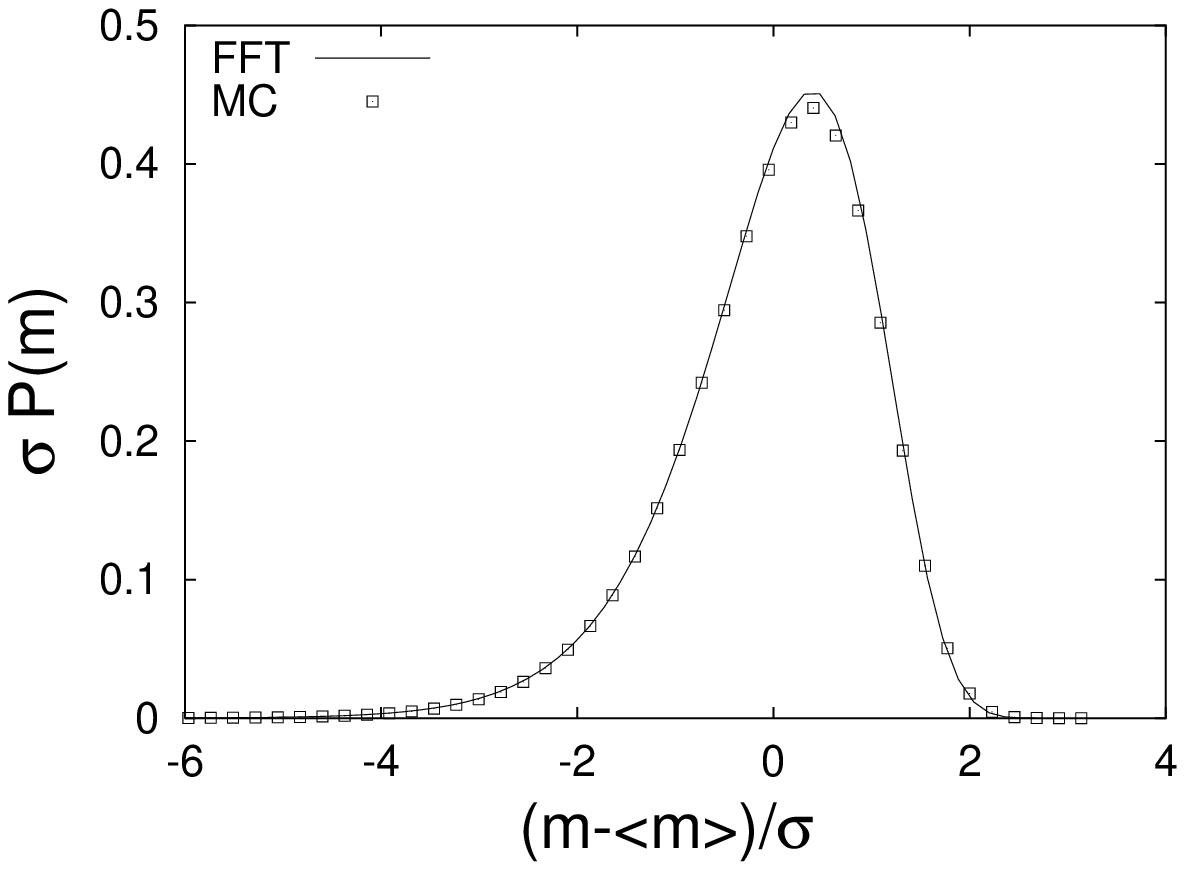,width=10cm}
\end{center}
\vspace{0.75cm}
\begin{center}
\epsfig{file=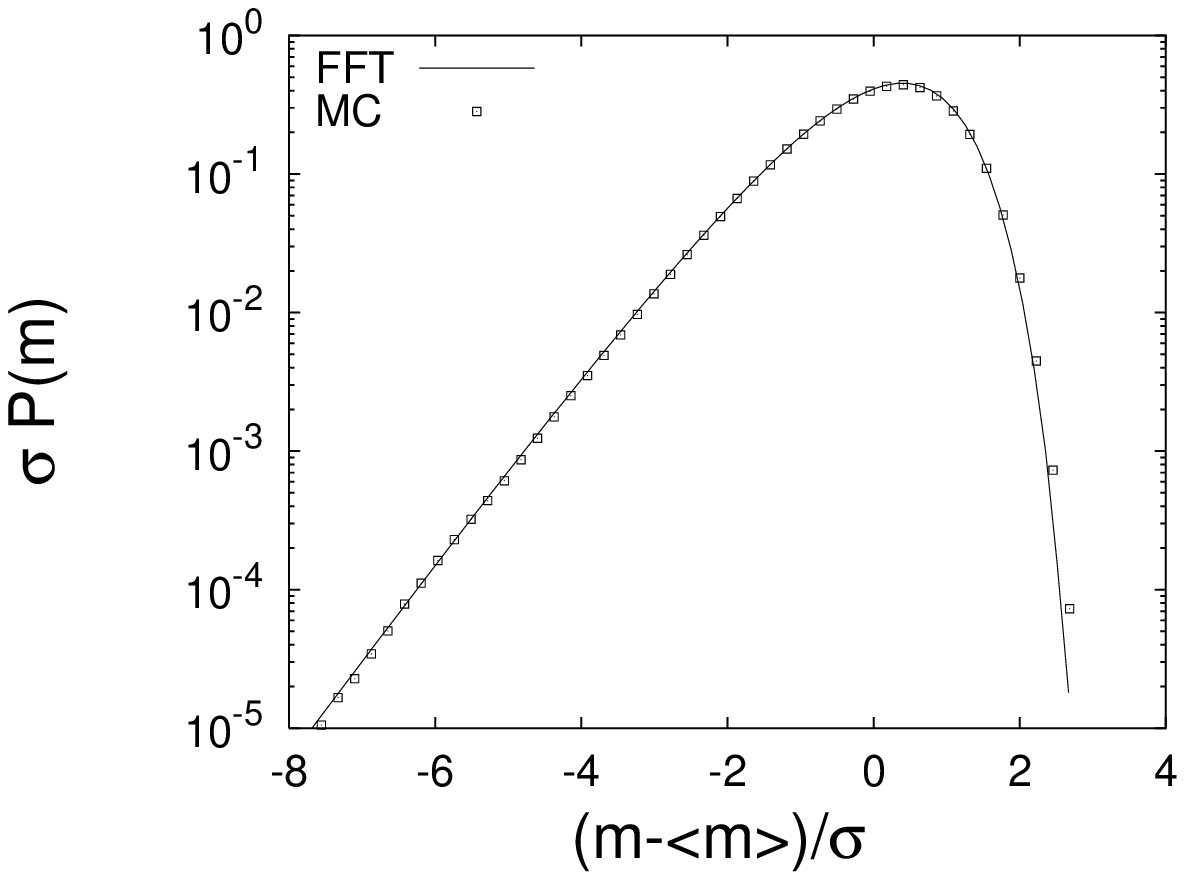,width=10cm}
\end{center}
\bigskip
\caption{The PDF, as obtained from a fast Fourier transform (FFT) of 
  equation~(\ref{eq10}), compared with MC simulation of a $2D$-XY model 
  at temperature $T=0.1$ of size $N=32^2$ (upper: natural scale; 
  lower: semi-log scale).}
\label{pdf_2d}
\end{figure}

The numerical integration of eqn.~(\ref{eq10}), performed with a fast
Fourier transform (FFT) algorithm~\cite{NR}, is shown in Fig.~\ref{pdf_2d},
where it is compared with Monte Carlo results for $T/J=0.1$ and $N=32^2$.
The theoretical curve is clearly
in extremely good agreement with the numerical data.
The curve is asymmetric, with what appears to be an exponential tail for
fluctuations below the mean, with a much more rapid fall off in amplitude,
for fluctuations above the mean.

\begin{figure}
\begin{center}
\epsfig{file=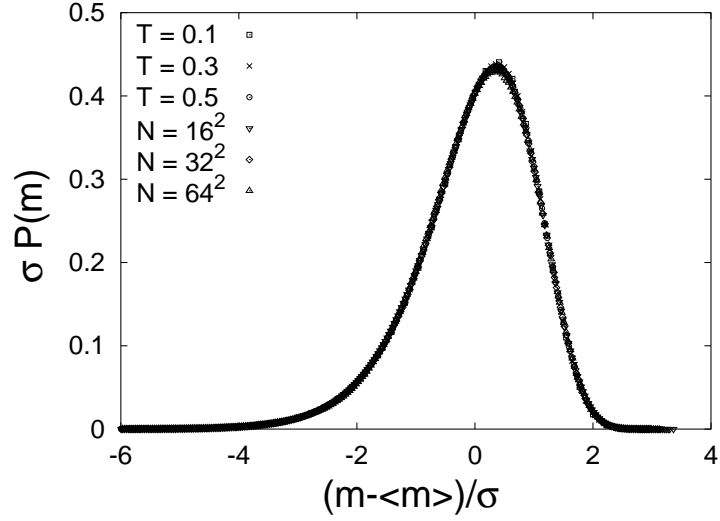,width=10cm}
\end{center}
\bigskip
\caption{The PDF  for fluctuations in dimension $D=2$ from 
  MC and MD simulations. The first set of data corresponds 
  to canonical MC simulation for a system of size $N=32^2$ 
  at temperature $T=0.1,\,0.3,\,0.5$.
  The second set of data corresponds to microcanonical MD 
  simulation at temperature $T \simeq 0.7$ and size 
  $N=16^2,\,32^2,\,64^2$.}
\label{mcmd}
\end{figure}

\begin{figure}
\begin{center}
\epsfig{file=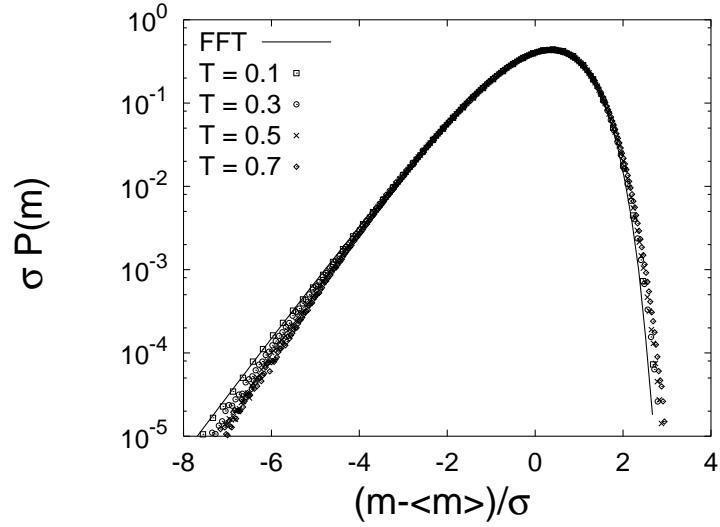,width=10cm}
\end{center}
\bigskip
\caption{The PDF, as obtained from a fast Fourier transform (FFT) of 
  equation~(\ref{eq10}), in dimension $D=2$ compared with Monte Carlo 
  results for a system of size $N=32^2$ at temperature 
  $T=0.1,\,0.3,\,0.5,\,0.7$.}
\label{MC_2D}
\end{figure}

\begin{figure}
\begin{center}
\epsfig{file=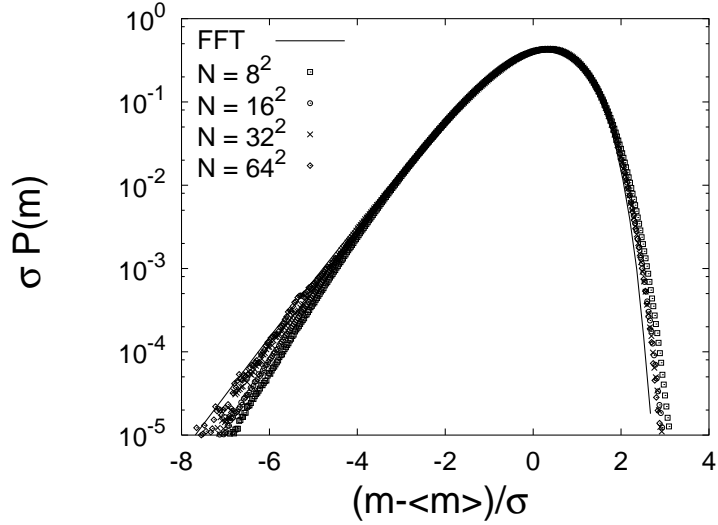,width=10cm}
\end{center}
\bigskip
\caption{The PDF, as obtained from a fast Fourier transform (FFT) of 
  equation~(\ref{eq10}), in dimension $D=2$ compared with MD results 
  for a system of size $N=8^2,\, 16^2,\, 32^2,\, 64^2$, at 
  temperature $T\simeq 0.7$. }
\label{MD_2D}
\end{figure}

In Fig.~\ref{mcmd} we show the PDF for fluctuations in $m$ obtained
from MC simulation for fixed system size and varying temperature, as well
as MD for fixed temperature and different system sizes. 
The result of ref.~\cite{bra.98} and section~\ref{sec.pdf_2} of this paper 
is that, for the harmonic Hamiltonian, eqn~(\ref{eq2}), $\Pi(\theta)$ is 
independent of both system size and temperature, while we have explicitly 
tested this result against the PDF generated for the full Hamiltonian, 
eqn.~(\ref{eq1}).
Qualitative agreement is clearly excellent, independently of the ensemble 
used, but there are small systematic deviations in the tails, when observed
on a logarithmic scales~\cite{pal.00}, as shown in 
Figs.~\ref{MC_2D} and~\ref{MD_2D}. 
We can only expect agreement between the analytic result and simulation in 
the range of temperature sufficiently below $T_{\rm KTB}$ such that vortex 
pairs do not influence the PDF~\cite{arc.97}.  
Even in the absence of vortices one must expect small variations from our
theoretical result for small system sizes that stem from the utilisation
of~(\ref{eq1}) rather than~(\ref{eq2}).
In a renormalisation group treatment the non-linearities of 
Hamiltonian~(\ref{eq1}) scale away on changing the length scale
and the Hamiltonian is replaced by an effective harmonic Hamiltonian 
at higher temperature~\cite{jos.77}. 
For example, at $T/J = 0.7$, for $L=32$, we find $\langle m \rangle =0.76$ 
from simulation, while eqn.~(\ref{eq2b}) gives $\langle m \rangle  = 0.81$. 
The effective coupling constant can be calculated by expanding the cosine 
and approximating the nonlinear terms using a Hartree 
approximation~\cite{vil.75}.
Renormalisation of the non-linearities introduces a microscopic length scale 
$a'$ which gives small corrections when compared with the calculated PDF.
However, this length scale is fixed by the temperature and the corrections 
should scale away as the ratio $a'/L \rightarrow 0$. 
This  scenario is confirmed  in Fig.~\ref{MD_2D}, where
data are shown at $T/J=0.7$, for
$L = 8,\;16,\; 32,\;64$ and compared with the theoretical curve. 
Deviations from the theoretical result are observed for $L=8$ and $L=16$
but the PDF clearly approaches the predicted scale independence for the 
larger system sizes.

%%%%%%%%%%%%%%%%%%%    FIGURE     %%%%%%%%%%%%%%%%%%%%%%
%%%%%%%%%%%%%%%%%%%%%%%%%%%%%%%%%%%%%%%%%%%%%%%%%%%%%%%%

Near  $T_{\rm KTB}$ vortices influence the PDF, however the vortex population 
decreases exponentially  moving away from $T_{\rm KTB}$~\cite{jen.91} and they
only make their presence felt within the physical domain in a small band of 
temperatures near the transition. 
In this regime the data do not fit on the universal 
curve~\cite{arc.97,sel.00,pal.00} but a detailed discussion of this point is 
outside the scope of this paper.

%%%%%%%%%%%%%%%%%%%%    FIGURE     %%%%%%%%%%%%%%%%%%%%%%
%%%%%%%%%%%%%%%%%%%%%%%%%%%%%%%%%%%%%%%%%%%%%%%%%%%%%%%%%

%%%%%%%%%%%%%%%%%%%%%%     FIGURE     %%%%%%%%%%%%%%%%%%%%%%%
%%%%%%%%%%%%%%%%%%%%%%%%%%%%%%%%%%%%%%%%%%%%%%%%%%%%%%%%%%%%%

%%%%%%%%%%%%%%%%%%%%%%     FIGURE     %%%%%%%%%%%%%%%%%%%%%%%
%%%%%%%%%%%%%%%%%%%%%%%%%%%%%%%%%%%%%%%%%%%%%%%%%%%%%%%%%%%%%

%%%%%%%%%%%%%%%%%%%%%%%%%%%%%%%%%%%%%%%%%%%%%%%%%%%%%%
\subsection{$P(m)$ for the Linearised Order Parameter}
\label{subsec.pdf_lop}
%%%%%%%%%%%%%%%%%%%%%%%%%%%%%%%%%%%%%%%%%%%%%%%%%%%%%%

As eqn.~(\ref{eq10}) is independent of temperature, one should be able
to obtain it at low temperature where the magnetisation is approximately
\be\label{m-lin}
m & = & 1-{1\over{2N}}\sum_i (\theta_i- \overline{\theta})^2.
\ee
In fact, using this expression one can arrive at (\ref{eq10}) in a more
straightforward manner. What is perhaps surprising is that the
calculation, using (\ref{m-lin}) is valid for all temperatures below 
$T_{\rm KTB}$, even for temperatures where  (\ref{eq-m}) and (\ref{m-lin}) 
represent different physical quantities.

Using the Hamiltonian (\ref{eq2}), we have
\be\nn
P(m) =  \frac{1}{Z}\int_{-\infty}^{\infty}\frac{dx}{2\pi}
\int \prod_i d\theta_i \exp\left\{
ix\left[ m-1+\frac{1}{2N}\sum_i \theta_i^2 \right]
-\frac{1}{2\tau}
\sum_{i,j}\theta_i G_{ij}^{-1}\theta_j\right\}
\ee
where $G_{ij}^{-1}$ is the inverse Green's function operator connecting 
sites $i$ and $j$ with non-zero elements for $i$ and $j$ nearest 
neighbours~\cite{ber.71}, and $Z=\left(\det G^{-1}/2\pi\tau\right)$ 
is the partition function.

It is easy to integrate the Gaussian integral  by transforming into
reciprocal space.
Defining the trace Tr of any function of $G$ as the sum for
${\bf q}\neq 0$ of the same function of $G({\bf q})$
and and using
$\langle m \rangle=1-\tau {\rm Tr}G/2N$ we find
\be\label{eq11}
P(m)=\int_{-\infty}^{\infty}\frac{dx}{2\pi}
\exp\left[ ix\left (m-\langle m \rangle \right )-ix\frac{\tau}{2}{\rm Tr} 
G/N -\frac{1}{2}{\rm Tr}\log \left ({\bf 1}-ix\tau G/N\right )\right ].
\ee
We can now use the fact that $\sigma=\sqrt{g_2/2}\tau$ in this approximation,
to transform (\ref{eq11}) into a dimensionless and universal form
\be\label{eq12}
\Pi(\theta)&=&\int_{-\infty}^{\infty}\sqrt{\frac{g_2}{2}}\frac{dx}{2\pi}
\exp\left [ix\theta\sqrt{\frac{g_2}{2}}
-i\frac{x}{2}\tr G/N-\frac{1}{2}\tr\log\left ({\bf 1}-ixG/N\right)
\right ]
\\ \nn
&=&\int_{-\infty}^{\infty}\sqrt{\frac{g_2}{2}}\frac{dx}{2\pi}
\exp\left [i \Phi(x)\right ]
\ee
which is the same expression as (\ref{eq9},\ref{eq10}), once we separate
the real and imaginary parts of the integrand.
This demonstration  proves that the only relevant graphs are those with only 
one loop, the others being zero in the thermodynamic limit.

Within this linear approximation the mean magnetisation $\langle m \rangle$ 
and the standard deviation $\sigma$ do not scale in the same way with system 
size: while 
$\langle m \rangle = 1- (T/8\pi J) \log(CN)$, $\sigma=\sqrt{g_2/2}\tau$ 
is a temperature dependent constant. This exact result can be verified by 
applying eqn.~(\ref{moments}) to eqn.~(\ref{eq11}) and calculating 
$\langle m \rangle $ and $\langle m^2 \rangle $ directly.
The fact that we find the same universal function for the two calculations,
when written in the form (\ref{scale-bhp}) shows explicitly that the
hyperscaling result, $\sigma/\langle m \rangle \sim O(1)$ is not a necessary
condition for non-Gaussian data collapse.
Rather, it seems that hyperscaling is a consequence, in these circumstances,
of the correct definition of $m$ as an order parameter on the interval $[0,1]$.

The Gaussian limit of the $2D$-XY model is identical to the Edwards-Wilkinson
model of interface growth and the linear approximation for the order parameter
is related to the square of the interface width $m = 1- w^2$.
The PDF for $w^2$ has been studied in one~\cite{fol.94} and two~\cite{rac.94} 
dimensions together with extensions to the EW model, including 
non-linearity~\cite{der.99,pra.00}.
All models give non-Gaussian PDF's with the same qualitative features as
Fig.~\ref{pdf_2d}.
These models provide an important microscopic link between equilibrium and
non-equilibrium systems and suggest that a formalism could exist that
incorporates the statistical features that we have observed to be shared,
at a global level, between such different systems.

%%%%%%%%%%%%%%%%%%%%%%%%%%%%%%%%%%%%%%%%%%%%%%%%%%%%%%%%%%%%%%%
\subsection{Asymptotes of $\Pi(\theta)$ for Large Fluctuations}
\label{subsec.pdf_asymptotes}
%%%%%%%%%%%%%%%%%%%%%%%%%%%%%%%%%%%%%%%%%%%%%%%%%%%%%%%%%%%%%%%

As a first step towards an analytic form for $\Pi(\theta)$
one can approximate (\ref{eq10}) beyond the Gaussian approximation
by retaining only the elements $(g_2,g_3)$. In this case,
the solution is proportional to the Airy function
\be
\Pi(\theta) \propto \exp\left (-\frac{1}{6\alpha}\theta\right )
{\rm Ai}\left[\frac{1}{(3\alpha)^{1/3}12\alpha}-\frac{1}{(3\alpha)^{1/3}}
\theta\right],
\ee
where $\alpha=2^{3/2}g_3/3g_2^{3/2}\simeq 0.296876$. The $g_3$ term assures
that it is not symmetric on reversing the sign of $\theta$. We  find that
the approximation reproduces qualitatively the apparent exponential
behaviour for  $\theta \ll -1$:
\be
\Pi(\theta)\sim
\left\{ 2\sqrt{\pi}
\left[ \frac{1}{(3\alpha)^{1/3}12\alpha}-\frac{1}{(3\alpha)^{1/3}} \theta 
\right]^{1/4}
\right\}^{-1} \exp\left\{
-\frac{1}{6\alpha}\theta-\frac{2}{3}\left [
\frac{1}{(3\alpha)^{1/3}12\alpha}-\frac{1}{(3\alpha)^{1/3}}
\theta \right ]^{3/2}\right\}
\ee
However the approximation does not allow us to extract the asymptote above
the mean, as for $\theta > 0$ the Airy function develops oscillations.

\begin{figure}
\begin{center}
\epsfig{file=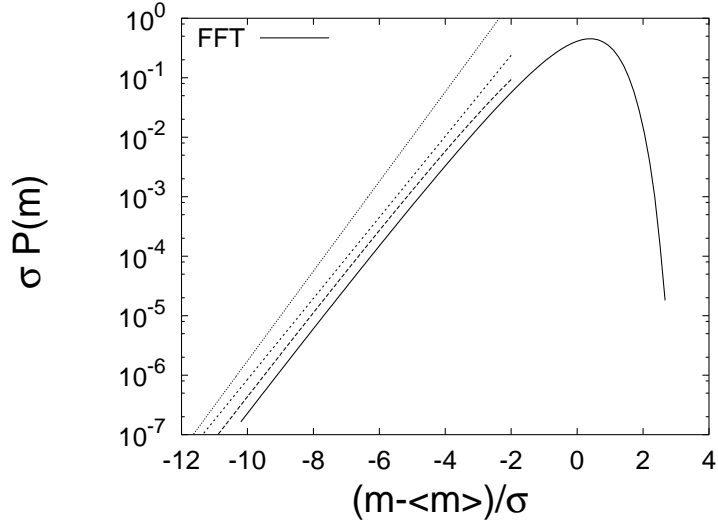,width=10cm}
\end{center}
\bigskip
\caption{Comparison of the tail of the PDF with the exact asymptote 
  (long dashed), eqn.~(\ref{eq16}), the true exponential tail of 
  slope $4\pi^2\sqrt{g_2/2}\simeq 1.736$ (dotted), and an effective
  exponential tail of slope $\alpha = 1.56867\dots$ (short dashed).
  The curves are displaced from each other for clarity.}
\label{tails}
\end{figure}

A more fruitful approach is to look at the saddle points of the  integrand
(\ref{eq10}), from which one can extract both asymptotes.
If $\theta \ll -1$, an expansion near $x=0$ is not very satisfactory and we
must rather seek the solution for the extrema of the whole integrand,
$\partial\Phi(x)/\partial x=0$. We find:
\be
%\nn
\sqrt{\frac{g_2}{2}}\theta
%&=&\frac{\partial}{\partial x}
%\left (\frac{1}{2}\tr
%\left \{ xG/N-\arctan [xG/N]\right \}
%-\frac{i}{4}\tr
%\log\left (1+x^2G^2/N^2\right )
%\right )
%\\
\label{eq13}
=\frac{1}{2}\tr \frac{G^3}{N^3}\frac{x^2}{1+x^2G^2/N^2}-
\frac{i}{2}\tr\frac{G^2}{N^2}\frac{x}{1+x^2G^2/N^2}.
\ee
 If $\theta$ is negative and $x$ real, the real part of the
second term is always positive and there is no solution
to this equation. 
We therefore seek a solution for $x$ pure complex, $x=iy$.
In this case, eqn.~(\ref{eq13}) becomes
\be\label{eq14}
\sqrt{\frac{g_2}{2}}\theta=\frac{1}{2}\tr
\frac{G^2}{N^2}\frac{y}{1+yG/N}=\varphi(y)
\ee
The function $\varphi$ has simple poles at
$y=-4\pi^2,-8\pi^2,-32\pi^2,\ldots$
and its asymptotic value near the first pole $y_0=-4\pi^2$ is
$\varphi(y)\sim -2/(y-y_0)$.
The extremum of the integrand satisfies the condition
$y^*\simeq y_0-2\sqrt{2/g_2}/\theta>y_0$, for $|\theta|$ large and we can
deform the real path of the integration so that it passes through the
extremum on the imaginary axis.
Near the extremum, we can expand the integrand up to second order in
$y-y^*$ and perform a Gaussian integration:
\be\label{eq15}
\Pi(\theta)\simeq
\int_{-\infty}^{\infty}\sqrt{\frac{g_2}{2}}\frac{dx}{2\pi}
\exp\left [ i\Phi(iy^*)+i\frac{1}{2}(x-iy^*)^2\Phi''(iy^*)\right ]
\ee
We finally find that the asymptotic value of the distribution
varies as
\be\label{eq16}
\Pi(\theta)\propto |\theta|\exp \left(4\pi^2\sqrt{\frac{g_2}{2}}\theta\right)
\ee
We have superimposed the asymptotic result (\ref{eq16}) and the full
 numerical integration for $N=101^2$ of (\ref{eq10}) in Fig.~\ref{tails}.
The amplitude of eqn.~(\ref{eq16}) is chosen so that the curves are slightly
displaced, to allow comparison of the slopes. The asymptotic solution
is in excellent agreement even for $\theta$ values where the PDF shows
a distinct deviation away from exponential behaviour and only fails for
$\theta > -2$. 
Further out in the tail, in the range $-10 < \theta < -4$,
$\log(\Pi)$ is approximately linear. However the value of the slope
is not the argument of the exponential in (\ref{eq16}),
 $4\pi^2\sqrt{g_2/2}\simeq 1.736$. The logarithmic corrections given by the
the term $|\theta|$ are significant over the whole of this range, but
the curvature is so small that the data can be fitted to an effective
exponential $\Pi(\theta) \sim \exp(\alpha \theta)$, with
$\alpha = 1.56867 \dots$.
The data only approaches true exponential behaviour for $\theta < -30$,
which is completely outside any imaginable physical range.  Strictly
speaking it is therefore more correct to speak of pseudo-exponential,
 $x\exp(\alpha x)$, for the asymptote below the mean.

%%%%%%%%%%%%%%%%%%%%%%%%     FIGURE     %%%%%%%%%%%%%%%%%%%%%%%%%%
%%%%%%%%%%%%%%%%%%%%%%%%%%%%%%%%%%%%%%%%%%%%%%%%%%%%%%%%%%%%%%%%%%

For large and positive $\theta$ a solution of eqn.~(\ref{eq14})
exists for large and positive $y$. A reasonable approximation is to
replace $G$ by $1/q^2$ and perform the integration
\be\label{large-as}
\varphi(y)
&\sim&\frac{1}{2}\int_{q=2\pi/\sqrt{N}}^{2\pi}\frac{Nd^2q}{4\pi^2}
\frac{1}{N^2q^4}\frac{y}{1+y/Nq^2}
\\ \nn
&\sim&\frac{1}{4\pi}\int_{2\pi/\sqrt{y}}^{\infty}
\frac{dq}{q(1+q^2)}\sim \frac{1}{8\pi}\log y
\ee
A more precise computation gives $\varphi=\log(y)/8\pi+\hat a +1/2y
+\cdots$
where $\hat a$ is a numerical constant which can be computed exactly.
An analytical study (see App.~\ref{app.2}) gives
\be\label{eq17}
\hat a= \frac{1}{24}+\frac{\gamma}{4\pi}-\frac{1}{4\pi}\log(4\pi)
-\frac{1}{2\pi}\log\prod_{k=1}^{\infty}\left[1-\exp(-2\pi k)\right]
=-0.11351444337\ldots
\ee
For large $\theta$, the saddle point of the integrand is therefore located
at $y^*=\exp 8\pi(-\hat a+\sqrt{\frac{g_2}{2}}\theta)$, and the asymptotic
value for $\Pi$ follows from a Gaussian integration of~(\ref{eq15}):
\be\label{eq18}
\Pi(\theta)\propto \exp \left[
-\frac{1}{8\pi} {\rm e}^{8\pi\left( \sqrt{\frac{g_2}{2}}\theta-\hat a\right)}
+8\pi\sqrt{\frac{g_2}{2}}\theta \right]
\ee
Comparing the asymptote with the full curve we again find that
the true asymptote only fits accurately outside the physical
domain, although the data is clearly consistent with a very rapid
fall off in the PDF for $\theta$ above the mean.

%%%%%%%%%%%%%%%%%%%%%%%%%%%%%%%%%%%%%%%%%%%%%%%%%%%%%%%%%%%%%%%%%%%%%%%%%%%%%%
\section{Fitting to known functional forms}
\label{sec.fitting}
%%%%%%%%%%%%%%%%%%%%%%%%%%%%%%%%%%%%%%%%%%%%%%%%%%%%%%%%%%%%%%%%%%%%%%%%%%%%%%

The obvious question now arises: is the PDF generated by the characteristic
function~(\ref{eq10}) of known functional form? 
We do not have a definitive answer to this question, as we are not able to 
transform~(\ref{eq10}) analytically.
In the absence of an answer, we test the PDF against three skewed  functions, 
shown in eqn.~(\ref{fit}), which describe statistics 
in different physical situations. 
These are: a modified Gumbel function, characteristic of problems where 
extreme values dominate the sum over many contributions;
a log-normal distribution, characteristic of statistically independent
multiplicative processes and a $\chi^2$ distribution which describes
the PDF of a quantity made up of a finite number of positive definite
microscopic variables. 
The analysis is the same in all three cases, but is only shown in detail for 
the modified Gumbel function: each curve has 4 parameters, but once the value 
of the first is chosen the others are fixed by normalisation and the 
constraints $\langle \theta \rangle  = 0$, $\langle \theta^2 \rangle  = 1$. 
The family of one parameter curves are Fourier transformed and the first 
four terms in a Taylor expansion are set equal to those for the generating 
function, which fixes the value of the free parameter. 
The method takes into account the skewness of the curve but not the kurtosis
and its accuracy is  ultimately limited.    
The goodness of fit can be measured by comparing the ratio of higher order 
terms of the expansion of the test and generating functions.
For an exact solution all higher ratios would be equal to unity, while for 
a poor fit they diverge rapidly from this value. 
Other functions could be tested in the same way and an exact solution may 
well exist in the statistics literature, unknown to us.

The method described above is quite similar to that due to 
Pearson~\cite{pea.1892}, who realised a century ago that, in all practical 
situations, knowledge of the first four moments of a distribution is 
sufficient to generate a curve, fitting any set of data 
points~\cite{elephant}. 
Pearson developed a phenomenological differential equation containing the 
numerical values of the moments, whose solution gives the fitting function. 
A Pearson analysis is performed on the calculated PDF at the end of the 
section.

%%%%%%%%%%%%%%%%%%%%%%%%%%%%%%%%%%%%%%%%%%%%%%%%
\subsection{The Generalised Gumbel Distribution}
\label{subsec.gumbel}
%%%%%%%%%%%%%%%%%%%%%%%%%%%%%%%%%%%%%%%%%%%%%%%%

The asymptotes (\ref{eq16}) and \ref{eq18}) are of the same general form 
as those for Gumbel's first asymptote distribution from the theory of
extremal statistics~\cite{gum.58}:
defining $z$ to be the $a^{th}$ largest value from a set of $z_i,\, i=1,N$ 
random numbers taken from a generator $f(z)$; the PDF for $z$  is
\begin{equation}\label{gumbel-1}
g_a(z) = \frac{a^a\alpha_a}{\Gamma (a)}
\exp\left\{-a\left[\alpha_a(z-u_a)+{\rm e}^{-\alpha_a(z-u_a)} 
\right] \right\}\,.
\end{equation}
$\Gamma(a)$ is the gamma function; $u_a$ is the value
of $z$ such that $a$ of the $N$ random numbers are greater than
$z$. $F(z)$ is the probability of having $a$ of the values less
than $z$, such that
 $F(u_a) = 1-a/N$. $\alpha_a$ is referred to as the intensity:
$\alpha_a= (N/a)f(u_a)$.
In conventional statistics, $a$ would of course be an integer.
However, in what follows we are going to see an irrational number 
appearing.

The function (\ref{gumbel-1}) has an exponential tail for fluctuations towards
large values of $z$, the opposite of the PDF, in Fig.~\ref{pdf_2d}. 
We therefore make a change of variables 
$m_z= 1-z,\; \theta_z= (m_z-\langle m_z \rangle )/\sigma_z$
which makes a mirror reflection of eqn. (\ref{gumbel-1}). 
Within the linear approximation for the order parameter this corresponds to 
the relevant variable being the sum of the spin wave amplitudes
$z \rightarrow (1/2N)\sum_i
(\theta_i-\overline{\theta})^2$~\cite{bra.00,fuckup}.
Changing variables we find
\begin{eqnarray}\label{gumbel}
\sigma_z \Pi_G(\theta_z) & =&  w {\rm e}^{a b(\theta_z-s) 
                              -a {\rm e}^{b(\theta_z -s)}} \nonumber \\
b & = & \alpha_a \sigma_z \nonumber \\
s & = & (1-\langle m_z \rangle -u_a)/\sigma_z \nonumber \\
w & = & {a^a\alpha_a\over{\Gamma (a)}} \sigma_z.
\end{eqnarray}
Eqn.~(\ref{gumbel}) is also the distribution for the $a^{th}$ smallest 
random number from the set $z_i$.
After some algebra one can show that
\begin{eqnarray}
b &= & 
\sqrt{ {1\over{\Gamma (a)}} {\partial^2 \Gamma (a)\over{\partial a^2}}- 
\left[ {1\over{\Gamma (a)}} {\partial \Gamma (a)\over{\partial a}}
\right]^2 } \nonumber \\
s & = &  {1\over{b}} 
\left[\log (a) - {1\over{\Gamma (a)}} {\partial \Gamma(a)\over{\partial a}} 
\right] \,.
\end{eqnarray}
Now re-writing (\ref{eq16}) and (\ref{eq18}), one finds
\begin{eqnarray}\label{asymp}
\Pi(\theta)&\propto & \left\{
\begin{array}{lll}
|\theta|\exp \left(\frac{\pi}{2} b \theta \right) \,, 
                    && \theta  \ll 0 \\ \\
\exp \left[ -\frac{\pi}{2}{\rm e}^{b (\theta-s)}+c\theta \right] \,,
                    && \theta \gg 0 \,,
\end{array}
\right.
\end{eqnarray}
with $b=8\pi\sqrt{g_2/2}\simeq 1.105$, $s= 0.745$ and $c= b$.
These asymptotes differ
only slightly from those for a generalised Gumbel function with $a=\pi/2$:
firstly through the term $|\theta|$ for fluctuations below the mean
and secondly through the term $\exp(c\theta)$ above the mean:
the coefficient $c=(\pi/2) b$ for the modified Gumbel function, while $c=b$
for the true asymptote.
These differences are enough to ensure that the modified Gumbel's equation
is not an exact solution to eqn. (\ref{eq9}), however the comparison is
so close that it is tempting  try to get a good fit to (\ref{eq10}) by
solving for the constants $a$, $b$, $s$ and $w$.

Fourier transforming (\ref{gumbel}) gives
\be
\Pi_G(\theta)=\int_{-\infty}^{\infty}\frac{dx}{2\pi}
\frac{w}{b}\exp \left (
ix\theta -isx+i\frac{x}{b}\log a-a\log a \right )
\Gamma \left(a-i\frac{x}{b} \right) =
\int_{-\infty}^{\infty}\frac{dx}{2\pi}\exp\left[i\Phi_G(x) \right] \,.
\ee
We can compare $\Phi_G(x)$ with $\Phi(x\sqrt{2/g_2})$, assuming that the two 
Fourier transforms are nearly equal. 
The four constants should be calculated by minimising the difference between
the two functions. 
To do this we can set the first four coefficients of the Taylor expansion of 
these functions equal. 
For $\Phi_G(x)$ we have:
\be\nn
\Phi_G(x)&=&ia\log a-i\log(w/b)-i\log\Gamma(a)
+\left[-s-\Psi(a)/b+\log(a)/b\right] x
+\frac{i}{2b^2}\Psi'(a)x^2
\\ \label{eq20}
&+&\frac{1}{6b^3}\Psi''(a)x^3
-\frac{i}{24b^4}\Psi'''(a)x^4-\frac{1}{120b^5}\Psi^{(4)}(a)x^5+\cdots
\ee
where $\Psi(z)$ is the digamma function $\Gamma'(z)/\Gamma(z)$.
For $\Phi$ we have:
\be\label{eq21}
\Phi(x\sqrt{2/g_2})=
\frac{i}{2}x^2-\frac{\sqrt{2}g_3}{3g_2^{3/2}}x^3-i\frac{g_4}{2g_2^2}x^4
+\frac{2\sqrt{2}g_5}{5g_2^{5/2}}x^5+\cdots
\ee
We therefore find that the four constants satisfy the relations:
\be\nn
\frac{b}{w}&=&\frac{\Gamma(a)}{a^a},\;\; sb=\log a-\Psi(a), \\ 
\label{eq22}
b^2 &=& \Psi'(a),\;\; b^3g_3\left(\frac{2}{g_2}\right )^{3/2}=-\Psi''(a)
\ee
The first three equations arise
from the constraints of normalisation of the distribution, while the last
expresses these constraints in terms of $g_2$ and $g_3$.
The equations can be solved numerically. We find
\begin{eqnarray}\label{eq23}
a &=& 1.5806801,\;\; b = 0.9339355 \nn \\
s &=& 0.3731792,\;\; w = 2.1602858
\end{eqnarray}
The constants  $b$ and $s$  calculated in this way are shifted slightly
from the values extracted from the asymptotes, but $a$ is close to our
very appealing first try
$\pi/2$. Taking this value and calculating
the constants $b$, $s$ and $w$ from normalisation one finds:
\begin{eqnarray}\label{eq23b}
a &=& \pi/2,\;\; b = 0.938 \nn \\
s &=& 0.374,\;\; w = 2.14,
\end{eqnarray}
in very satisfying agreement with the first method of calculation.

Given this solution, we can compute the coefficient ratio
for the higher order terms in (\ref{eq20}) and (\ref{eq21}):
\be
\frac{1}{\Phi_G^{(4)}(0)}
\left. \frac{\partial^4\Phi(x\sqrt{2/g_2})}{\partial x^4}
\right |_{x=0}&=&
\frac{12g_4b^4}{g_2^2\Psi'''(a)}=0.9265029,
\\
\frac{1}{\Phi_G^{(5)}(0)}
\left. \frac{\partial^5\Phi(x\sqrt{2/g_2})}{\partial x^5}
\right|_{x=0}&=&
-\frac{48\sqrt{2}g_5b^5}{g_2^{5/2}\Psi^{(4)}(a)}=0.8267429
\ee
The ratio of coefficients clearly does diverges from unity, but is does so
slowly, indicating that the modified Gumbel function should be a good fit
to the curve over the physical range.
This is confirmed in Fig.~\ref{Gumbel} where we compare~(\ref{gumbel}),
using the values (\ref{eq23}), with  the exact result,
from eqn. (\ref{eq10}). On a natural scale the agreement is
remarkably good over the entire range, with the only visible deviation
coming around the maximum of the PDF, where the Gumbel curve is very
slightly lower. On a logarithmic scale there is excellent general agreement
over the whole of the plotted range, but a slight deviation can be observed
for probabilities below $10^{-3}$. For fluctuations below the mean the
deviation is because the true asymptotic behaviour is quasi-exponential,
$x\exp(-\alpha x)$ and has a slight curvature, as discussed in the previous
section. The results therefore confirm that, although the generalised Gumbel
function is an excellent approximation for the PDF (\ref{eq10}), it is
not an exact solution.

>From these results it is very tempting to take the generalised Gumbel 
function, with $a$ exactly $\pi/2$ as a working analytic expression
for the PDF. 
However the connection with extremal statistics remains an open 
question~\cite{cha.00}.
As discussed in section~\ref{sec.conclusion}, the spin wave 
Hamiltonian~(\ref{eq2}) is diagonalised in reciprocal space and the 
problem can be formulated in terms of a set of statistically independent 
variables. 
The PDF for extreme values of statistically independent variables can 
only follow three different asymptotic~\cite{gum.58,bur.99}, or limit 
functions as the thermodynamic limit is taken. 
The only possible limit functions from extremal statistics of the Gumbel 
form discussed here are for $a$ integer, with $a=1$ for the biggest or 
smallest values.

Chapman {\it et al.}~\cite{cha.00}, have recently argued that the PDF for 
global quantities in any system with identifiable excitations on scales 
up to the system size should be dominated by extreme values. 
They shown that the PDF of extreme values among $10^5$ Gaussian 
random number generators approximates to a Gumbel function
with $ a = \pi/2$. 
This is not one of the predicted asymptotes~\cite{bur.99}, and we suggest 
that  the deviation must be due to a very slow approach to the limit 
function with system size. 
It therefore does not seem to be a correct description of the $2D$-XY data 
as we do have a limit function which is well represented by 
eqn.~(\ref{gumbel}) with $a=\pi/2$.
However, if the results of~\cite{cha.00} are relevant for 
non-equilibrium phenomena such as turbulence and self-organised 
criticality it would suggest the interesting property that corrections
to the asymptotic forms, or limit functions,are a generic feature of these
systems.

%%%%%%%%%%%%%%%%%%%%%%%%%%%%%%%%%%%%%%%%%%%%%%%%
\subsection{Generalised log-normal Distribution}
\label{subsec.lognormal}
%%%%%%%%%%%%%%%%%%%%%%%%%%%%%%%%%%%%%%%%%%%%%%%%

The generalised log-normal distribution has the form
\begin{equation}
\Pi_L(\theta)=\frac{w}{\sqrt{2\pi\sigma_L^{2}}
(s-\theta)}\exp 
\left\{-\frac{1}{2\sigma_L^{2}} \left[\ln (s-\theta)-a \right]^{2}
 \right\} \,,
\end{equation}
with $w=1$. Following the same procedure as before, the generating function
$\Phi_L(x)$ can be developed as a power series
\begin{equation}\label{gen-L}
\Phi_L (x)=
x \left( \theta-s+{\rm e}^{a+\sigma_L^2/2} \right)
+ i\frac{x^{2}}{2}\Bigl({\rm e}^{2a+2\sigma_L^2}-{\rm e}^{2a+\sigma_L^2}\Bigr)-x^{3}
\Bigr(\frac{1}{6}{\rm e}^{3a+9\sigma_L^2/2}+\frac{1}{3}{\rm e}^{3a+3\sigma_L^2/2}-\frac{1}{2}{\rm e}^{3a+5\sigma_L^2/2}\Bigr) \,.
\label{qet}
\end{equation}
Comparing (\ref{gen-L}) with (\ref{eq21}) one finds the
following expressions for
$s,\,a$ and $\sigma_L$:
\be
s &=& {\rm e}^{a+\sigma_L^2/2} \nonumber \\
a &=& -\frac{1}{2}\ln\left( {\rm e}^{2\sigma_L^2}-{\rm e}^{\sigma_L^2}\right) 
          \nonumber \\
\frac{\sqrt{2}}{3}\frac{g_3}{g_2^{3/2}} &=& \frac{1}{6}{\rm e}^{3a}
\left( {\rm e}^{9\sigma_L^2/2}+2{\rm e}^{3\sigma_L^2/2}-
3{\rm e}^{5\sigma_L^2/2}\right).
\ee
Eliminating $a$ and $\sigma_L$ leads to a cubic equation for $s$ in terms of
$\alpha=\frac{(g_{2}/2)^{3/2}}{g_{3}} = 1/|\gamma|$:
\begin{equation}
s^{3}-3\alpha s^{2}-\alpha=0,
\label{car}
\end{equation}
which could be solved exactly. We have solved it numerically, verifying
that there exists one real and two complex roots. We find
\begin{center}
 $s=3.45981$, $a=1.20109$, $\sigma_L=0.28325$.
\end{center}

The function, with these parameters, is compared with the calculated PDF 
in Fig.~\ref{Gumbel}. 
The general quality of fit is again excellent over the plotted range, with 
very small systematic deviations occurring in the wings of the distribution. 
It does not have the correct asymptotes; either exponential on the left
or double exponential on the right, but as we have shown in the previous
section, the true asymptotic behaviour is only reached outside the plotted
regime, which explains why such a good fit can be achieved.

We have not, for the moment been able to develop a physical reasoning
associated with the log-normal function and the origin, $s=3.4...$ 
although related to $\gamma$, seems rather arbitrary, but we do not exclude 
an explanation in terms of random multiplicative processes.

Note that log-normal distribution does appear in surface dynamics. 
Namely, starting with a flat interface as an initial condition, 
the short-time limit of the $D=1$ Edwards-Wilkinson dynamics yields 
a log-normal distribution for the interface width~\cite{Antal}.

\subsection{Generalised $\chi^2$ Distribution}
\label{subsec.chisquare}

\begin{figure}
\begin{center}
\epsfig{file=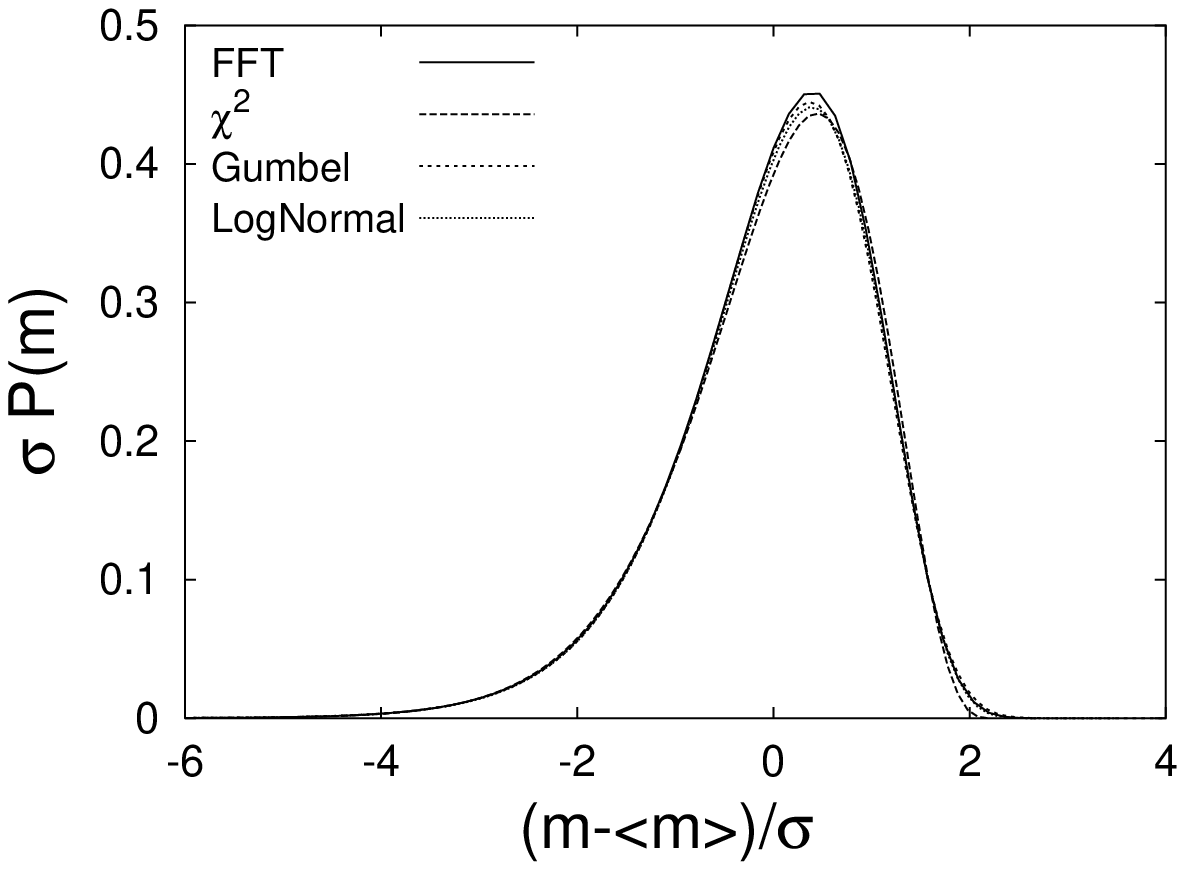,width=10cm}
\end{center}
\vspace{0.75cm}
\begin{center}
\epsfig{file=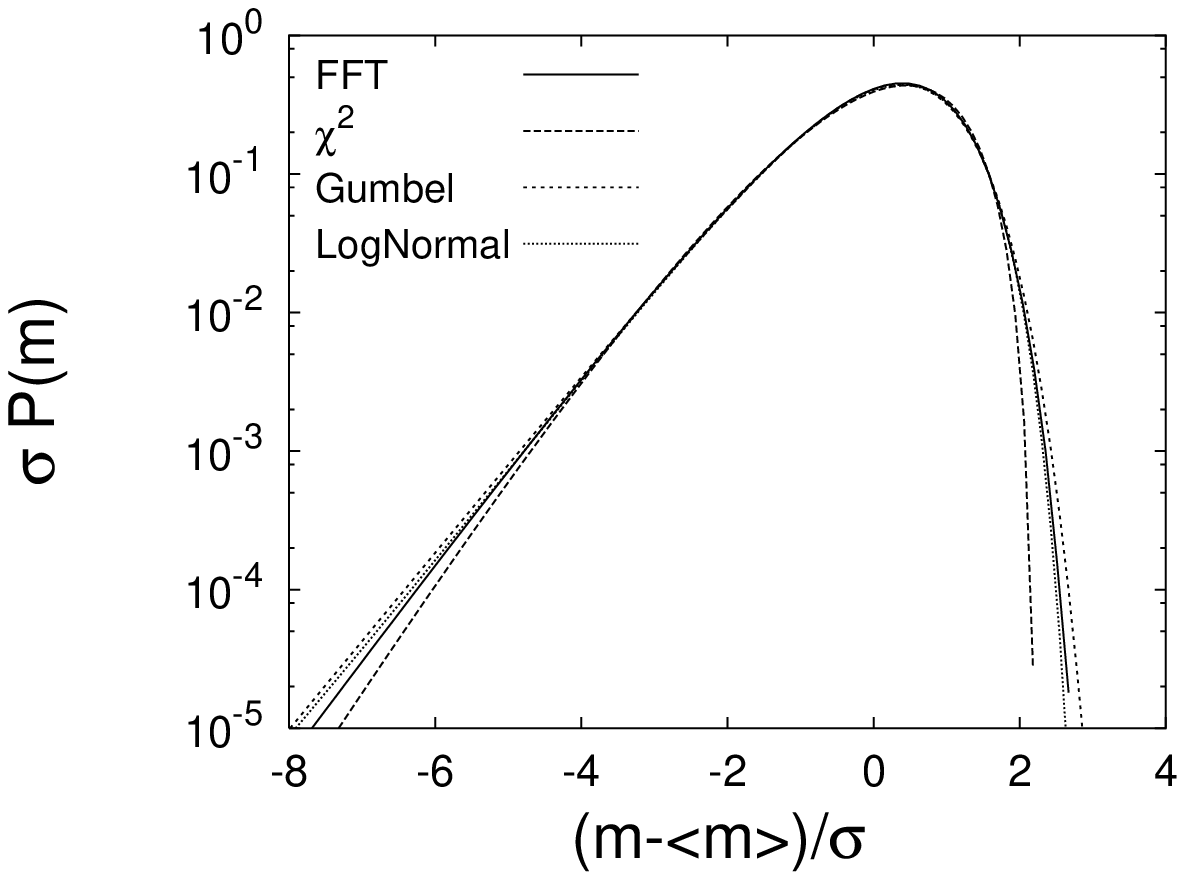,width=10cm}
\end{center}
\bigskip
\caption{The PDF compared with the generalised Gumbel, 
  log-normal and $\chi^2$ functions described in the text.}
\label{Gumbel}
\end{figure}

The $\chi^2$ distribution for $\nu$ statistically independent degrees
of freedom has the form
\begin{equation}
\Pi_{\chi}(\theta)=w(s-\theta)^{\nu/2-1}{\rm e}^{-a(s-\theta)},
\end{equation}
with
\be
w &=&\frac{a^{\nu/2}}{\Gamma(\nu/2)} \nonumber \\
\nu&=& 2a^2.
\label{varc}
\ee
As in the case of the Gumbel function, the generating function can
be found in closed form:
\begin{equation}
\Phi_{\chi}(x)=x(\theta-s)+i\frac{\nu}{2}\ln(1-ix/a),
\end{equation}
whose development up to  $4^{th}$ order in $x$ leads to
\begin{equation}
\Phi_{\chi}(x)=x(\theta-s)+\frac{\nu}{2a}x+i\frac{\nu}{4a^2}x^2
-\frac{\nu}{6a^3}x^3-i\frac{\nu}{8a^{4}}x^4+ O(x^5) +\dots
\end{equation}
This series can again can be compared  with~(\ref{eq21}) to give
\be
s&=&\frac{\nu}{2a}\nonumber \\
a&=&\sqrt\frac{\nu}{2}=s \nonumber \\
\nu&=&\frac{g_2^3}{g_3^2} = \frac{8}{\gamma^2},
\ee
with numerical values
\begin{center}
$\nu=10.07155$, $a=2.24405$ , $s=a$ , $w=2.31233$.
\end{center}
Comparing the function, shown in Fig.~\ref{Gumbel} with these parameters
with the calculated curve, there is reasonably good  agreement but this time 
deviation can be seen when plotted both on real and logarithmic scale. 
On the logarithmic scale the deviation is stronger than for the other 
fitting functions.

One can see that describing the correlated system as a finite number of 
degrees of freedom is a reasonably good approximation. 
It is an appealing concept and the calculation  yields  a system size 
independent number which depends uniquely on the skewness:
$\nu =g_2^3/g_3^2= 8/\gamma^2$. 
If $\gamma$ developed towards zero, then $\nu$ would diverge and the $\chi^2$ 
interpretation would be consistent with a Gaussian distribution.       
However, quantitatively it is not correct and the true description is a 
many body one~\cite{por.00}. 
The difference between the two curves can be quantified by considering the 
ratio of the $4^{th}$ order terms:
\be
\Phi_{\chi}(x)^{(4)}&=&-i\frac{1}{2\nu} \nonumber \\
\Phi(x)^{(4)}&=&-i\frac{g_{4}}{2g_{2}} ,
\ee
so that $\frac{\Phi(x)^{(4)}}{\Phi_{\chi}(x)^{(4)}}\sim 0.0238$
which is very far from $1$.

%%%%%%%%%%%%%%%%%%%%%%     FIGURE     %%%%%%%%%%%%%%%%%%%%%%%%%
%%%%%%%%%%%%%%%%%%%%%%%%%%%%%%%%%%%%%%%%%%%%%%%%%%%%%%%%%%%%%%%

\subsection{Pearson's Curve}
\label{subsec.pearson}

Pearson \cite{pea.1892,sma.58} described an ingenious method of deriving a
functional form for a PDF to fit experimental data, given the first four
moments of the latter. He considered the differential equation
\begin{equation}
\frac{d\ln{y}}{dx} =  - \frac{x + b}{b_0 + b_1 x + b x^2}
\end{equation}
\noindent
and showed that if $y$ is a distribution then the parameters $b$, $b_0$,
$b_1$ are specific functions of the first four moments. The expression can
then be integrated to give (within a normalisation factor) an approximate
functional form for the PDF, which by definition has the same principal
moments as the data to be fitted. The success of Pearson's approach relies
on the observation that PDFs with the same moments are approximately
coincident over the range of a few standard deviations, which is exactly
the range of experimental interest. In the present case the mean is zero
and the standard deviation is set to unity, so the shape of the curve
depends only on the skewness, $\gamma$, and kurtosis, $\kappa$.

\begin{figure}
\begin{center}
\epsfig{file=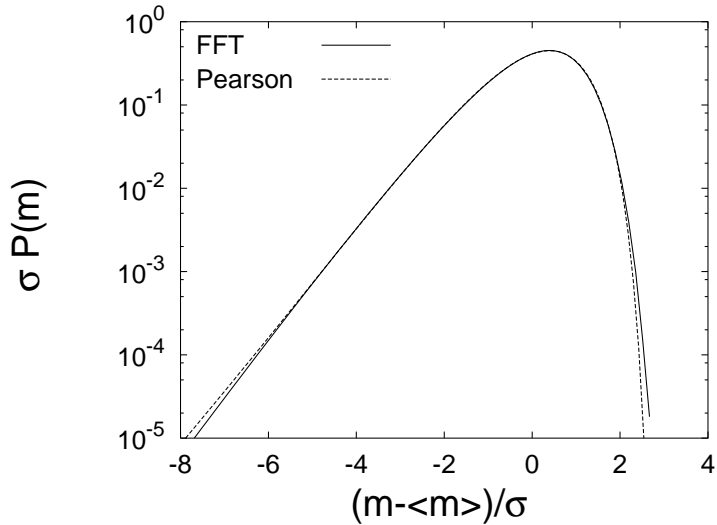,width=10cm}
\end{center}
\bigskip
\caption{The PDF compared with the fit obtained with the Pearson 
  method described in the text.}
\label{Pearson}
\end{figure}

%Using the values $g_2 = 3.8667;10^{-3}$,  $g_3 = 7.5719;10^{-5}$,
% $g_4 = 1.7626;10^{-6}$, 
We find 
$\gamma  = g_3 (2/g_2)^{3/2} = -0.8907$, 
and 
$\kappa = 3 + 3 g_4 (2/g_2)^2 = 4.415$, 
which gives the following solution:
\begin{equation} \label{P-curve}
y = y_0 \frac{(\beta - \xi)^q}{(\alpha - \xi)^p}
\end{equation}
in which $\xi = x - 0.39723$, $\beta = 2.4787$, $\alpha = 11.430$,
$q = 10.249$, $p = 47.267$, $y_0 = \exp{(105.02)}$. 
Equating $y(x) = \Pi(\theta)$, the fit to the exact 
expression~(Fig.~\ref{Pearson}) is good between  $x = -6$ and $x = 2$, 
but the very large numbers involved in eqn.~(\ref{P-curve}) suggest that 
this functional form has no physical significance. 
From this analysis we can conclude that data collapse observed in 
refs.~\cite{bra.00} should be interpreted as meaning that the third 
and fourth moments scale with $\sigma$ and $L$ in the same way as
they do in the critical $2D$-XY model.

%%%%%%%%%%%%%%%%%%%%%%%%     FIGURE     %%%%%%%%%%%%%%%%%%%%%%%%%
%%%%%%%%%%%%%%%%%%%%%%%%%%%%%%%%%%%%%%%%%%%%%%%%%%%%%%%%%%%%%%%%

%%%%%%%%%%%%%%%%%%%%%%%%%%%%%%%%%%%%%%%%%%%%%%%%%%%%%%%%%%%%%%%%%%%%%%%%%
\section{Distribution in the $D$-dimensional Gaussian model}
\label{sec.pdf_D}
%%%%%%%%%%%%%%%%%%%%%%%%%%%%%%%%%%%%%%%%%%%%%%%%%%%%%%%%%%%%%%%%%%%%%%%%%

\begin{figure}
\begin{center}
\epsfig{file=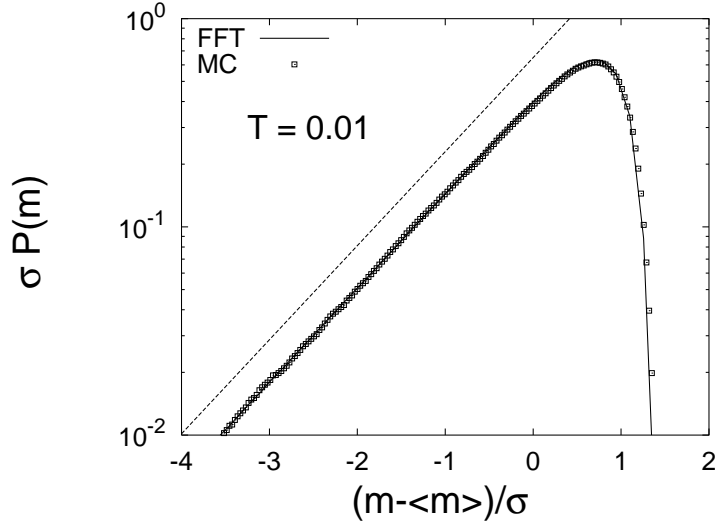,width=10cm}
\end{center}
\bigskip
\caption{The PDF in one dimension ($N=128$) at temperature 
  $T/J < 12/N$. The dashed line (with slope $\simeq 1.04$) 
  is the exponential asymptote for the low temperature 
  approximation given by eqn.~(\ref{eq31}) and is shifted 
  with respect to the main curve for clarity.}
\label{pdf_1d}
\end{figure}

In this section, we study the asymptotics of the distribution function in 
general dimension $D$. It is straightforward to generalise the development
from eqn.~(\ref{eq9}) to eqn.~(\ref{eq10}) for arbitrary dimension by 
redefining $G({\bf q})$ for dimension $D$ and summing over a $D$ dimensional
Brillouin zone.
The generalised expression (\ref{eq10}) can then be numerically transformed
to give $\Pi(\theta)$.
The results for $D=1$ and $D=3$ are shown in Figs.~\ref{pdf_1d} 
and~\ref{pdf_3d}, where they are compared with data from Monte Carlo and 
molecular dynamics simulations.
There is again excellent agreement showing that eqn.~(\ref{eq10}) is 
essentially  exact in the low temperature regime, where the 
Hamiltonian~(\ref{eq2}) is  valid.
At higher temperatures the full Hamiltonian~(\ref{eq1}) generates vortex 
structures, the eqn.~(\ref{eq5}) is no longer valid and the derivation of 
eqn.~(\ref{eq10}) breaks down.
Within the low temperature approximation there are three regimes:
$D<2$, $2<D<4$, and $D \ge 4$, in addition to the special case $D=2$.
The different regimes can be seen from a dimensional analysis of $g_1$
and $g_2$.
As deviation from Gaussian behaviour is due to the abnormal influence of the
integral scale in the form of infrared divergences,
we approximate replacing $G$
by $1/q^2$ and re-calculate the $g_k$ by performing
integrals over the Brillouin zone between $2\pi/N^{1/D}$ and $2\pi$.
This procedure gives the correct qualitative behaviour, but there is a
difference between the discrete sums and the integrals over the Brillouin
zone, even in the thermodynamic limit (see App.~\ref{app.2}).
The correct qualitative behaviour is
\be\label{eq27}
g_1 &\simeq & \left\{
\begin{array}{llr}
C_{1,D}N^{(2-D)/D}\,, & D<2  \\ \\
A_1\log N + B_1 \,,   & D=2   \\ \\
C_{1,D} \,,  & D>2
\end{array}
\right.
\ee
and
\be
\label{eq271}
g_2 &\simeq & \left\{
\begin{array}{llr}
C_{2,D}N^{2(2-D)/D} \,, & D<4  \\ \\
\left (A_2\log N+B_2\right )/N \,, & D=4  \\ \\
C_{2,D}/N \,, & D>4.
\end{array}
\right.
\ee
The lower and upper critical dimensions, $D=2$ and $D=4$, are marked by
the logarithmic divergence of $g_1$ and $g_2$ respectively.

Using the linearised order parameter (\ref{m-lin})
we find  for $D<2$ that $g_1$ diverges as
a power of $N$ giving $\langle m \rangle = 1- \tau C_{1,D} N^{(2-D)/D}$,
which is a poor approximation  for a thermodynamic quantity bounded
on the interval $[0,1]$.
Once outside this restricted low temperature region,
$\tau \le 1/\left[C_{1,D}N^{(2-D)/D}\right]$,
both the linear approximation for the order parameter and the quadratic 
approximation for the Hamiltonian break down and there is a divergence in
the behaviour of the PDF, as calculated from~(\ref{eq10}) and as simulated 
numerically.
The system is, of course, disordered at all temperatures, so that the correct
$\langle m \rangle$ and $\sigma$ both vary as $1/\sqrt{N}$ and the PDF for the
vector order parameter is a two dimensional Gaussian function centered on
${\bf m} = 0$.
The PDF for $m$, as defined in (\ref{eq-m}), is
$P(m) \sim m \exp{(-m^2/2\sigma^2)}$, analogous to a Maxwellian
distribution of molecular speeds, and the thermodynamic system satisfies
the central limit theorem (see App.~\ref{app.CLT}).
As we have already seen, for $D=2$ the situation is different,
as there is a large region of temperature where the quadratic Hamiltonian
correctly describes the physics even though eqn. (\ref{m-lin})
is not a good approximation. In this regime of temperature, the
PDF $\Pi(\theta)$, for parameters (\ref{m-lin}) and (\ref{eq-m})
are however identical.

For dimension $D>2$, the low temperature expansion for the order
parameter gives consistent results for all $N$, as long range order is
stable and $\langle m \rangle \sim 1 -C_{1,D}\tau$ is well defined.
Above $D=4$, our results agree with mean field theory ($D=\infty$)
where all sites are connected together.
Here, $\langle m \rangle\simeq 1 - \tau/4$ and $\sigma\simeq\tau/2\sqrt{2N}$,
and for large but finite $N$, the universal function $\Pi$ is simply a
Gaussian
\be
\Pi(\theta)=\frac{1}{\sqrt{2\pi}}\exp \left ( -\frac{1}{2}\theta^2
\right ),
\ee
which corresponds to the central limit theorem for a collection of $N$
independent oscillators, each of expectation value $\langle m \rangle $ and
standard
deviation $\tau/2\sqrt{2N}$.

%%%%%%%%%%%%%%%%%%%%%%%%     FIGURE     %%%%%%%%%%%%%%%%%%%%%%%%%%
\begin{figure}
\begin{center}
\epsfig{file=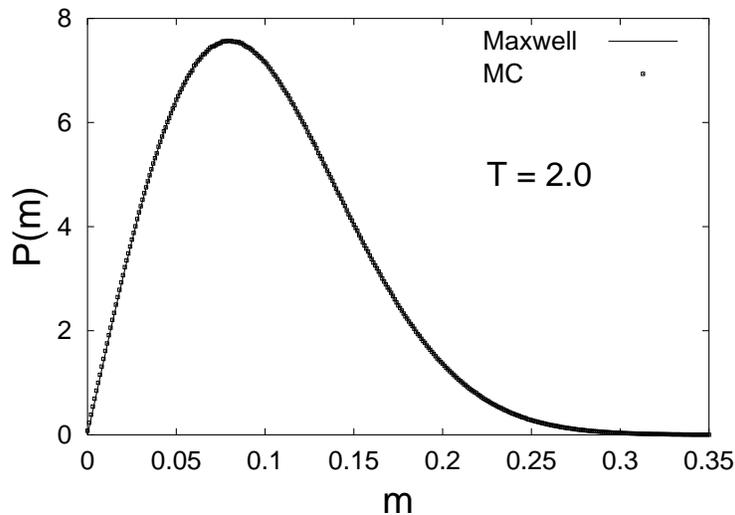,width=10cm}
\end{center}
\bigskip
\caption{The PDF in one dimension ($N=128$)  at temperature $T/J > 12/N$.
The continuous line is Maxwell speeds distribution of an ideal gas.
}
\label{Maxwell}
\end{figure}
%%%%%%%%%%%%%%%%%%%%%%%%%%%%%%%%%%%%%%%%%%%%%%%%%%%%%%%%%%%%%%%%%%

%%%%%%%%%%%%%%%%%%%%%%     FIGURE     %%%%%%%%%%%%%%%%%%%%%%%
%%%%%%%%%%%%%%%%%%%%%%%%%%%%%%%%%%%%%%%%%%%%%%%%%%%%%%%%%%%%

%%%%%%%%%%%%%%%%%%%%%%%%%%%%%%%%%%%%%%%%%%%%%%%%%
\subsection{Low Temperature Calculation in $D=1$}
\label{subsec.pdf_1}
%%%%%%%%%%%%%%%%%%%%%%%%%%%%%%%%%%%%%%%%%%%%%%%%

If the low temperature calculation for $D<2$ is not terribly pertinent for the
thermodynamic system, it it highly  relevant for the interface problem in the
context of the EW model~\cite{rac.94,fol.94,pli.94} and is  exactly
solvable in $D=1$~\cite{fol.94}.
In this case, computing the different $g_k$, we find
\be\nn
g_1=N/12,\;\; g_2=N^2/720, \ldots, \;\;
g_p=\frac{2 \, \zeta(2p)}{(2\pi)^{2p}}N^p \qquad p \gg 1  \,,
\ee
where $\zeta(k)=\sum_{i=1}^{\infty} i^{-k}$ is the Riemann zeta 
function~\cite{olv.97}.
The expectation value of the magnetisation and standard deviation are
\be\nn
\langle m \rangle &=&\exp\left(-\frac{\tau N}{24} \right)
\simeq  1-\frac{\tau N}{24},
\\ \nn
\sigma^2=\left (
\frac{1}{N}\sum_{{\bf r}}\cosh{\tau G_R({\bf r})}-1\right )\langle m \rangle^2
&=&\left (\int_0^1\cosh \tau N(x^2-x+1/6)dx -1\right )\langle m \rangle^2
\sim \tau^2 N^2 \,,
\ee
which means that the ratio $\langle m \rangle /\sigma$ scales as $1/N$,
although for the parameters of the interface problem
$\langle w^2 \rangle /\sigma_{w^2}\sim~O(1)$.
We evaluate the universal distribution $\Pi$, using the
general eqn.~(\ref{eq12}) with $G$ defined for the $D=1$.
After some algebra, we find for $\Pi(\theta)$
\be
\Pi(\theta)&=&\int\frac{dx}{2\pi}\exp
\left[ ix 
  \left( \theta-\frac{\sqrt{360}}{12} \right) -\sum_{k=1}^{\infty} \log 
  \left( 1-\frac{ix\sqrt{360}}{2 \pi^2 k^2} \right )
\right]
\nonumber \\ 
&=&\int\frac{dx}{2\pi}\exp
\left[ ix \left( \theta-\frac{\sqrt{360}}{12}\right) -
\log \left(\frac{\sin \sqrt{ix\sqrt{360}/2}} 
{\sqrt{ix\sqrt{360}/2}} \right) \right] \nonumber \\
&=&  
\int\frac{dx}{2\pi}\exp\left[i\Phi(x)\right]
\label{eq25}
\ee
This expression is related directly to the function $\tilde\Phi$ of
eqn.~(11) in ref.~\cite{fol.94}:
$\Pi(\theta)=\tilde\Phi(2-24\theta/\sqrt{360})$. 
The method used in~\cite{rac.94,fol.94,pli.94} is based on path integration, 
but the results are the same as our  saddle point method, used to compute 
the asymptotics.
Setting $x=iy$, the extrema of $\Phi$ satisfy the equation
\be\label{eq26}
\theta-\frac{\sqrt{360}}{12}&=&-\frac{\sqrt{360}}{2\pi^2}
\sum_{k=1}^{\infty}\frac{1}{k^2+y\sqrt{360}/{2\pi^2}}
\\ \nn
&=&-\frac{\sqrt{360}}{2\pi^2}\left (
-\frac{1}{2y\sqrt{360}/2\pi^2}+\frac{\pi}{2\sqrt{y\sqrt{360}/2\pi^2}}
\coth \pi\sqrt{y\sqrt{360}/2\pi^2}\right ).
\ee
For $\theta\ll -1$, $y$ is close to the first pole $-2\pi^2/\sqrt{360}$ of
the right hand side of ~\ref{eq26}), which is similar to the $2D$ case
(\ref{eq14}) except that the $1D$ extrema function is easier to evaluate.
Performing the saddle point computation, we find that $\Pi$ behaves
asymptotically as
\be\label{eq31}
\Pi(\theta)\propto \exp\left[ 2 \pi^2\theta/\sqrt{360}\right]
\ee
which is the same as \cite{fol.94}.
The asymptote, (\ref{eq31}), is drawn on Fig.~(\ref{pdf_1d} b), where 
it can be compared with the full calculation and with simulation. 
The exponential tail is extremely well defined and the predicted slope 
is clearly correct.

In the regime of fluctuations above the mean, for $y \gg 1$, $\theta$ is
close to the constant $\sqrt{360}/12$, and no extrema exist for $\theta$
beyond this value. 
In this case, $y\simeq \sqrt{360}/[8(\sqrt{360}/12-\theta)^2]$, and the 
saddle point approximation leads to the following asymptotic value for 
$\Pi$ near this upper limit
\be
\Pi\left (\theta \right )\propto
\left (\sqrt{360}/12-\theta \right )^{-5/2}
\exp \left ( - \frac{1}{8}\sqrt{360}/(\sqrt{360}/12-
\theta) \right ),
\ee
which is the same result as \cite{fol.94}. We refer the reader to
ref.s~\cite{rac.94,fol.94,pli.94} for the precise coefficients
in both asymptotic limits.

In conclusion, we find that for $\theta\ll -1$, the universal distribution 
again has an exponential tail, while for large fluctuations above the mean 
the PDF shoots to zero as $\exp[-3\theta_0/2 (\theta - \theta_0)]$, with 
$\theta_0 =\sqrt{360}/12$.
This upper limit corresponds to the constraint that $m \le 1$.

It is worth pointing out in some detail here that the exponential tail in 
the one-dimensional problem is not the result of critical fluctuations. 
The small deviations in angle $(\theta_i-\theta_j)$  constitute a random 
walk with $w \sim \sqrt{1-m}$  being the radius of gyration, which scales 
correctly as the square root of the walk length, $L$. 
The $1D$ linear order parameter, or interface problem is therefore nothing 
more than a simple random walk~\cite{fol.94}, but despite this the PDF is, 
as shown in Fig.~\ref{pdf_1d}: 
a standard result for such a walk is that the mean radius of gyration is 
proportional to the mean end to end distance, $S$ of the walk. 
It is easily shown that the PDF $P(S)$ is Gaussian~\cite{edwards}.
Changing variable from $S$ to $X=S^2$ one finds 
$P(X) \sim X^{-1/2} \exp(-X/X_0)$; a trivial distribution with an exponential
tail. 
The PDF for $w^2$ has the same exponential tail, but does not show the 
essential singularity at $w^2= 0$ ($m=1$) and we conclude the rather 
surprising property of a random walk, that the PDF for the radius of gyration 
and for the end to end distance are not the same. 
The origin of this difference is that the average angle $\overline{\theta}$,
corresponding to the center of mass of an equivalent random  walk, fluctuates 
with $L$ in the same way as the radius of gyration itself and  this lack of 
self averaging removes the essential singularity from the PDF at $w^2=0$.

%%%%%%%%%%%%%%%%%%%%%%%%%%%%%%%%%%%%%%%%%%%%%%%%%%%%%%
\subsection{Asymptotic Solutions in General Dimension}
\label{subsec.pdf_3}
%%%%%%%%%%%%%%%%%%%%%%%%%%%%%%%%%%%%%%%%%%%%%%%%%%%%%

We first evaluate the asymptotic value of $\Pi$ for positive $\theta$ by
solving the saddle point of~(\ref{eq12}), rescaling the
variable $x\sqrt{g_2/2}\rightarrow x$ for convenience. For $D<2$,
the ratio $g_1/\sqrt{g_2}$ is independent of the system size
and, with $x=iy$, the equation to solve is
\be
\theta-\frac{g_1}{\sqrt{2g_2}}\propto- \int_{\C/N^{1/D}}^{\C}
\frac{Nq^{D-1}}{q^2N+y\sqrt{2/g_2}}dq
\ee
where Cst is a constant. By setting $N^{1/D}q/\sqrt{y}\rightarrow q$,
we find that, for large and positive $y$
\be
\label{eq28}
\frac{g_1}{\sqrt{2g_2}}-\theta \propto
y^{(D-2)/2} \int_{\C/\sqrt{y}}^{\C N^{1/D}/\sqrt{y}}\frac{q^{D-1}}{1+q^2}dq
\sim y^{(D-2)/2}\int_0^{\infty}\frac{q^{D-1}}{1+q^2}dq
\ee
which means that $\theta$ is close to the upper bound $g_1/\sqrt{2g_2}$.
Replacing the asymptotic value of $y$ for the extrema in the function
$\Phi$ (\ref{eq12}), we find that
\be
\log \Pi(\theta) \sim -\C \left (
\frac{g_1}{\sqrt{2g_2}}-\theta \right )^{D/(D-2)}+{\rm
Logarithm\;\;corrections},\;\;\theta\sim \frac{g_1}{\sqrt{2g_2}},\;\;
D<2.
\ee
The logarithmic corrections come partly from the Gaussian integration
around the saddle point and partly from other terms in~(\ref{eq28}) which
are not accurately evaluated within our approximation.
Note again that $D=2$ is a special case as, instead of~(\ref{eq28}) we have
a logarithmic divergence (see eqn.~(\ref{large-as}))
and subsequently a double exponential
fall in $\Pi$ for large $\theta$.
For the interval $2 < D < 4$ the ratio $g_1/\sqrt{g_2}$ and the integral
(\ref{eq28}) are no longer finite and so we look to eqn.~(\ref{eq14})
for the asymptotic behaviour:
\be\label{eq29}
\theta\propto \frac{1}{g_2}\int_{\C/N^{1/D}}^{\C}\frac{d^Dq}{Nq^4}\frac{y}{
1+y\sqrt{2}/(\sqrt{g_2}Nq^2)}.
\ee
By again setting $N^{1/D}q/\sqrt{y}\rightarrow q$ and using the fact that
$g_2N^{2(D-2)/D}$ is finite (\ref{eq27}), we arrive at
\be\label{eq30}
\theta\propto y^{(D-2)/2}\int_0^{\infty}\frac{q^{D-3}dq}{1+q^2},\;\;y\gg 1
\ee
The integral is convergent for $2<D<4$ and by replacing the value for $y$ 
in the saddle point approximation, we get the asymptotic form for $\Pi$, 
in the  limit of  large and positive $\theta$:
\be
\log \Pi(\theta)\sim -\C \theta^{D/(D-2)}+{\rm Logarithm\;\;corrections},
\;\;\theta\gg 1,\;\;2<D<4.
\ee%
In 3 dimensions, we therefore expect that the logarithm of the distribution
falls off like $\theta^3$, well above the mean. We have not tested this in
detail, but the PDF does fall off more slowly for $D=3$ than $D=2$, in 
qualitative agreement with the predictions here.
Finally, we note that throughout the range $2<D<4$ the universal PDF is 
non-Gaussian, but the hyperscaling relation is invalid:
$
\langle m \rangle / \sigma \sim g_1/ \sqrt{g_2} \sim N^{(D-2)/D}
$.

For $D>4$, $g_2$ decreases as $1/N$, consequently,~(\ref{eq30})
has to be modified. We find, instead of~(\ref{eq30}), that
\be\label{eq32}
\theta\propto
y^{(D-2)/2}N^{(4-D)/4}
\int_{N^{(D-4)/4D}/\sqrt{y}}^{N^{1/4}/\sqrt{y}}\frac{q^{D-3}dq}{1+q^2}
\sim y,\;\;N\gg 1
.
\ee
We can, in fact, replace the integrand inside the integral by $q^{D-5}dq$
since the integration domain is large,
from which we find that the saddle point is proportional to $\theta\gg 1$ 
and deduce that $\Pi$ is Gaussian on the right hand side of the curve. 
The same is true for $D=4$ despite the logarithmic divergence of $g_2$.

\begin{figure}
\begin{center}
\epsfig{file=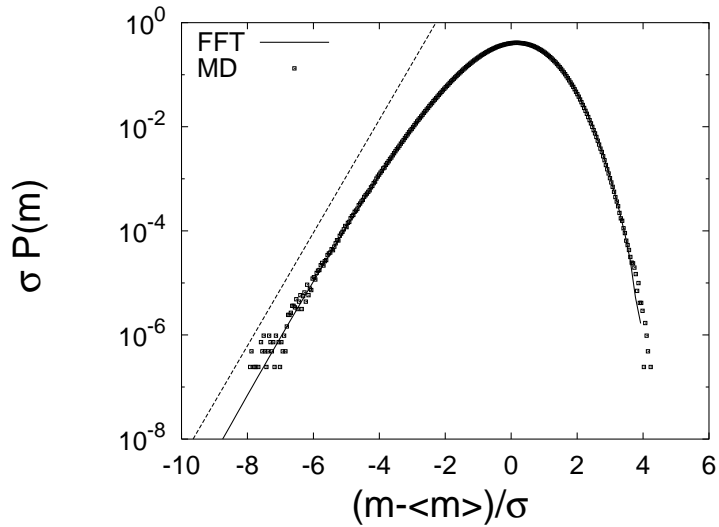,width=10cm}
\end{center}
\bigskip
\caption{The PDF in three dimensions for $N= 8^3$ and $T/J = 1.82$.
The dashed line (with slope $\simeq 2.5$) is the exponential asymptote
given by eqn.~(\ref{3d-as}) and is shifted with respect to the main curve
for clarity.}
\label{pdf_3d}
\end{figure}

In the opposite limit $\theta\ll -1$, for both  $D = 1$
and $D = 2$ the asymptotic value of the distribution falls down
exponentially~(\ref{eq16},\ref{eq31}). 
We would now like to evaluate this limit in general
dimensions. In both cases the coefficient of $\theta$ is related to
the value of $g_2$, i.e. $C_{2,D}$.
 Rewriting the eqn.~(\ref{eq14}) with discrete sums 
(see also App.~\ref{app.2}), we have
\be
\theta=\frac{N^{2(2-D)/D}}{16\pi^4g_2}
\sum_{m_i\ge 0}{}^{'}\frac{1}{(m_1^2+\cdots+m_D^2)}\frac{y}{(m_1^2+\cdots
+m_D^2)+y\sqrt{2/g_2}N^{(2-D)/D}/4\pi^2}
\ee
where the sum excludes $m_i=0$, $i=1,\, \ldots,\, D$. 
The saddle point equation has a solution $y$ which is the pole nearest the 
origin, $y=-4\pi^2\sqrt{g_2/2}N^{(D-2)/D}$, i.e. for sets of $\{m_i\}$ with
one element equal to 1, the others being zero. 
For $D<4$ and large $N$, this pole is finite since $g_2$ compensates 
$N^{2(D-2)/D}$, so that its value is simply $y=-4\pi^2\sqrt{C_{2,D}/2}$. 
Applying the saddle point integration, we find that the dominant term in 
the logarithm of $\Pi$ is, below the mean
\be\label{3d-as}
\log \Pi(\theta)\sim 4\pi^2\sqrt{\frac{C_{2,D}}{2}}\theta,\;\;\theta\ll -1,
\;\;D<4 \,,
\ee
and is linear in $\theta$ for every dimension below 4.
Included in Fig.~\ref{pdf_3d} for $D=3$ is a fit, on the left hand side of the 
form~(\ref{3d-as}), with $C_{2,3}$ calculated numerically.
There is again excellent agreement, which convincingly confirms the
presence of the exponential tail. 
In fact, true exponential behaviour is reached for smaller values of 
$\theta$ than for $D=2$.

For $D>4$, the value of this pole diverges like $N^{(D-4)/2D}$, and the
previous solution fails.
In fact, the solution~(\ref{eq32}) for positive $\theta$ and $y$ is
also valid for negative values, if $q^{D-1}dq/(q^2+1)$ is replaced
by $q^{D-1}dq/(q^2-1)$.
Since the integration domain is far from the
pole of the denominator, we can approximate the integrand in both cases by
$q^{D-5}dq$, and we get the same result as~(\ref{eq32}).
We therefore, finally conclude that $\Pi$ is also Gaussian on the left hand
side of the curve and the central limit theorem applies for $D > 4$.

%%%%%%%%%%%%%%%%%%%%%%%%     FIGURE     %%%%%%%%%%%%%%%%%%%%%%%%%%
%%%%%%%%%%%%%%%%%%%%%%%%%%%%%%%%%%%%%%%%%%%%%%%%%%%%%%%%%%%%%%%%%%

%%%%%%%%%%%%%%%%%%%%%%%%%%%%%%%%%%%%%%%%%%%%%%%%%%%%%%%%%%%%%%%%%%%%%%%%%%%%%
\section{Conclusion}
\label{sec.conclusion}
%%%%%%%%%%%%%%%%%%%%%%%%%%%%%%%%%%%%%%%%%%%%%%%%%%%%%%%%%%%%%%%%%%%%%%%%%%%%%

Probability functions with exponential, rather than Gaussian behaviour are 
a common feature of complex 
systems~\cite{bou.97,der.99,pra.00,fri.95,kad.99,har.99}.
For example, the PDF for velocity differences at microscopic scales, in fully 
developed turbulence show exponential tails~\cite{fri.95}.
This appears to be true in turbulence, not only for microscopic quantities, 
but also for global quantities; the energy injected into a closed turbulent 
flow being a very well controlled and documented 
example~\cite{lab.96,pin.99,aum.00}. 
Following these observations we have proposed that this is also a generic 
feature of complex systems~\cite{bra.98,bra.00}.
In this paper we have shown that, for the low temperature phase of the XY 
model, a critical system at equilibrium, analogous behaviour occurs when 
a few long wavelength and large amplitude modes make their presence felt 
in the global measure, which is typically a sum over $O(N)$ degrees of 
freedom.
The exponential tail can occur in three physically different situations. 
The first is in two-dimensions, when the system is critical and fluctuations 
occur over all length scales. The second is in one-dimension, when the system 
is not critical, but an exponential tail occurs for a particular global 
measure, relevant to problems of interface growth, whose moments are 
completely dominated by the integral scale.  
The third is in three-dimensions, also non-critical, where despite stable 
long range order, the large amplitude long wavelength  modes continue to 
make their presence felt.
The detailed form of the PDF in these three cases are quite different and 
easily discernible in experiment. 
In table 1 we show the evolution of the skewness and the kurtosis with 
spatial dimension. 
The deviations from the Gaussian limit are largest in one dimension, and
decrease continually to zero at $D = 4$. We propose that the difference 
in the form of the PDF could be used as an experimental signature of the 
underlying physics.

From the general evolution shown in table 1, one might expect a dependence 
on shape, with dimensional crossover as the length scale in one direction 
changes from microscopic to macroscopic.
This is indeed the case and for example, in two-dimensions, the skewness 
and kurtosis of the PDF calculated from eqn.~(\ref{eq10}) increases towards 
the values for $D =1$ if the ratio of lengths in the $x$ and $y$ directions, 
$L_x$ and $L_y$, are varied continuously from unity. 
It would be extremely interesting to establish if the same is true when the 
length scales are varied in turbulence experiments and numerically, in the 
models of self-organised criticality.

To see how the anisotropy of the PDF comes from the long wavelength 
excitations, we give an analysis in reciprocal space:
the Hamiltonian~(\ref{eq2}) is diagonalised
\be
H = \frac{J}{2} \sum_{{\bf q}>0} G({\bf q})^{-1} \Re \{\phi_{\bf q}\}^2,
\ee
where $\phi_{\bf q}$ is the discrete Fourier transform of $\theta_i$ and 
the sum is over the Brillouin zone~\cite{weg.67}, with the thermodynamic 
variable for each ${\bf q}$ taken as the real part of $\phi_{\bf q}$. 
Defining $m_{\bf q} = (1/2N) \Re \{\phi_{\bf q}\}^2$ the linear order 
parameter can be written $m = 1 - \sum_{\bf q} m_{\bf q}$, where the 
$m_{\bf q}$ are statistically independent variables with PDF
\be\label{micro}
P(m_{\bf q}) &=& \sqrt{\beta J q^2 N\over{4 \pi}} 
m_{\bf q}^{-1/2} \, {\rm e}^{-\beta J N q^2 m_{\bf q}} \,.
\ee
Here, as we are principally interested in the modes at small $q = |{\bf q}|$ 
we have, without loss of generality, approximated 
$G({\bf q})^{-1} \approx q^2$.
The PDF for $m$ is thus nothing more than the composite PDF for a set of
independent spin wave modes or an ``ideal gas'' of particles, whose only 
peculiarity is that the   mass   term varies as $q^{-2}$.
The Goldstone modes have wave vector $q = 2\pi/L$ and hence make 
contributions of $O(1)$ to $m$, while the modes on the zone edge, with 
$q = \pi$ have only microscopic amplitude.
This dispersion in amplitudes is the key to the unusual behaviour for $D=2$,
as it violates one of the conditions for the central limit theorem to apply 
to a sum of statistically independent variables: that the individual
amplitudes do not differ by too much. 
However it is not true that the Goldstone modes, by themselves give the 
complete PDF.
The mean value
$\left\langle \sum_q m_q \right\rangle \sim \int_{2\pi/L}^{\pi} q^{-2}n(q) dq$
where $n(q) \sim q^{D-1}$ is the density of states.
For $D=2$ both limits of the integral are required and a detailed calculation
gives $\langle \sum_q m_q \rangle  = ({\eta/4}) \log (CN)$, with $C=1.87$ and
with critical exponent $\eta = T/2\pi J$.
The anomalous term $\log N$ therefore reflects the fact that modes from
all over the Brillouin zone are relevant for $\langle m \rangle$ and
through eqn.~(\ref{exp}), for
the higher moments $\langle m^p \rangle$.

For $D = 1$ only the lower limit of integration is required, the upper limit 
can be set to $\infty$ and the constants $g_p$ are proportional to $N^p$. 
As a result the linear development of the order parameter in small 
angles,~(\ref{m-lin}), is a very poor approximation for the thermodynamic 
quantity defined on the interval $[0,1]$. 
The two expressions,~(\ref{eq-m}) and~(\ref{m-lin}),
describe different physical quantities. 
The former is directly related to the interface width in the 
Edwards-Wilkinson model of interface growth.  
The PDF for the full order parameter is consistent with  an 
uncorrelated system, that is, a paramagnet with two-dimensional order 
parameter. 
For the linear order parameter the PDF, shown in Fig.~\ref{pdf_1d}, does have 
an exponential tail, but this is not the result of critical fluctuations, 
it is the property of a simple random walk.
We remark further that dependence on a macroscopic length scale does not, in
itself, indicate critical behaviour.
Rather, critical behaviour is exemplified by the case $D=2$, where all length 
scale are important between the microscopic and macroscopic cut off.
    
$D=3$ represents the opposite of the one dimensional case: $\langle m \rangle$
is controlled by the upper limit of integration and the result is unchanged 
by setting the lower limit to zero. 
However, despite long range order being stable and the system not being 
critical in the low temperature phase, the exponential tail persists. 
This is related to temperature being a dangerously irrelevant 
variable~\cite{gol.92} near the zero temperature fixed point of a
 renormalisation group flow, between the lower and the upper critical 
dimension. 
The constant $g_2$ now falls to zero with system size but it does so more 
slowly than $1/N$ (see eqn.~(\ref{eq271})).
As a result of this slow decay, the ratio $g_p/g_2^{p/2},\;p > 2$ in 
eqn.~(\ref{eq9}) is independent of $N$ and the distribution is non-Gaussian,
despite $g_p$ and $g_2$ both being zero in the thermodynamic limit.
At low temperature the  magnetisation is finite, but the Goldstone mode
influences the PDF sufficiently to produce an exponential tail.
A physical consequence of this anomaly is that the
longitudinal  susceptibility
\be
\chi \sim {N\over{T}} 
\left( \langle m^2 \rangle - \langle m \rangle^2 \right)
\sim N^{(4-D)/D}
\ee
is weakly divergent throughout the ordered phase~\cite{arc.97,hol.40}.
This is true for all magnetic systems with Heisenberg or XY
symmetry. It could therefore be interesting to look for evidence of the
departure from Gaussian behaviour experimentally in
a non-critical three-dimensional system. Precision temperature control
would not be required, however, as the ratio $\sigma/ \langle m \rangle$ 
falls off as $1/N^{1/3}$, the divergence in
the susceptibility is very weak and this phenomenon
may be out of experimental reach.

\begin{center}
\begin{tabular}{|c|c|c|}\hline
$D$ & $\gamma$ & $\kappa$  \\ \hline
$\;\;\;1\;\;\;$ & $-1.807$ & $\;\;8.14\;\;$    \\ \hline
2 & $-0.891$ & $4.41$    \\ \hline
3 & $-0.354$ & $3.31$    \\ \hline
4 & $ 0$ & $3.0$    \\ \hline
\end{tabular}
\medskip

\medskip

{\small TABLE 1 Variation of skewness $\gamma$ and kurtosis $\kappa$ with
dimension $D$}
\end{center}

Returning finally to critical systems; we have been able to exploit a system 
interacting via a quadratic Hamiltonian at exactly the lower critical 
dimension. 
In this particular case one has access to a critical point, with the 
fluctuation dominated behaviour that this implies, while retaining the 
benefit of Gaussian integration over phase space. 
As a result, all critical behaviour can be calculated microscopically, without 
the need for either the renormalisation group or the scaling hypothesis. 
The only price one pays for this simplicity is a critical system with a single 
independent exponent and the scaling relations satisfied through weak scaling 
only.
In general, we believe that the analytic results that we have obtained are
useful for the understanding of finite-size scaling and for the 
interpretation of experimental observations from more complex correlated 
systems. 
The examples we have discussed~\cite{bra.98,bra.00} point towards
a behaviour analogous to criticality for an enclosed turbulent flow
and for models showing self-organised criticality.
However the detailed analysis presented here leaves many open questions
and more experiment
and simulation are clearly required if the generality and the limits
of this proposition are to be tested further.

%%%%%%%%%%%%%%%%%%%%%%%%%%%%%%%%%%%%%%%%%%%%%%%%%%%%%%%%%%%%%%%%%%%%%%%%%%%%
\acknowledgements{
This work was largely motivated by our collaboration with K.~Christensen, 
H.J.~Jensen, S.~Lise, J. L\'{o}pez and M.~Nicodemi from Imperial College 
London and we are particularly grateful to H.J.~Jensen and M.~Nicodemi 
for bringing the theory of extremal statistics to our attention. 
In addition we have greatly  benefitted from discussions with S.~Fauve, 
N.~Goldenfeld, J.~Harte, A.~Noullez, and Z.~R\'acz during the SIMU/CECAM
planning meeting ``Universal Statistics in Correlated Systems'', in Lyon, 
29-31 March 2000 and from subsequent discussions with L.~Berthier, 
E.~Leveque, S.~McNamara, P.~Pujol and again Z.~R\'acz in connection with 
the $1D$ problem.
It is our pleasure to thank all these people.

MS acknowlegdes a Marie Curie fellowship of the European Commission
(contract ERBFMBICT983561).    
The numerical work was supported by the P\^ole Scientifique de 
Mod\'elisation Num\'erique at the \'Ecole Normale Sup\'erieure 
de Lyon.  
}

%%%%%%%%%%%%%%%%%%%%%%%%%%%%%%%%%%%%%%%%%%%%%%%%%%%%%%%%%%%%%%%%%%%%%%%%%%%
\appendix
%%%%%%%%%%%%%%%%%%%%%%%%%%%%%%%%%%%%%%%%%%%%%%%%%%%%%%%%%%%%%%%%%%%%%%%%%%%

\setcounter{equation}{0}

%%%%%%%%%%%%%%%%%%%%%%%%%%%%%%%%%
\section{}\label{app.CLT}
%%%%%%%%%%%%%%%%%%%%%%%%%%%%%%%%%

\subsection*{Some comments on the central limit theorem in critical systems}

The central limit theorem is a powerful result of probability theory that
provides the foundation for statistical thermodynamics~\cite{khi.49}.
It states that the PDF of the sum $Z= \sum_{i=1}^N z_i$ of $N$ statistically 
independent variates $z_i$ tends, in the limit of large $N$ and for moderate 
values of the variate $Z$, to a Gaussian distribution. 
As well as the statistical independence of the $z_i$, another key criterion 
for the theorem to hold is that the $z_i$ are 
{\it individually negligible}~\cite{fel.71,gri.92,mis.64}. 
At a critical point, the first of these criteria is violated. 
The $2D$ XY model is of particular interest here as it is diagonalisable into 
statistically independent degrees of freedom and maps directly onto a 
problem where the second criterion is violated: the direct space
variables, that is, the spins $S_i$, are certainly individually negligible 
for large system size $N$, but are strongly correlated. 
On the other hand, when diagonalised in reciprocal space the spin wave 
variables are statistically independent, but are no longer all individually
negligible. 
In particular, the long wavelength modes make a significant impact on the 
fluctuations of the global measure; in this case the  linearised order 
parameter (\ref{m-lin}). 
The PDF for the full and the linear order parameters are identical, even 
when the quantities themselves differ, which makes it an ideal  system for 
the practical study of the breakdown of the central limit theorem.
A conventional critical system cannot, in general be
diagonalised in this way, as evidenced by the divergent specific heat.

Strictly speaking, the central limit theorem does not apply to the compound
variate $Z$, but rather to the normalised quantity 
$(Z - \langle Z \rangle)/N^{1/2}$.
This normalisation is essential for a reasonable PDF in the thermodynamic 
limit, as the standard deviation for fluctuations about the mean value 
$\langle Z \rangle$ scales with system size in the same way. 
If a normalisation factor $N^{1/2 + \rho}$, $\rho \ne 0$, is chosen then one 
obtains a distribution that is concentrated either at zero or 
infinity~\cite{cas.78}. 
We illustrate this with an example from statistical thermodynamics. 
The total energy $E$ of an ideal gas of $N$ molecules has a PDF of the 
form $P(E) \sim E^{3N/2 - 1} \exp{(-\beta E)}$. 
It is straightforward to confirm that $P(E)$ tends to a delta function in 
the thermodynamic limit, while $P(E/N^{1/2})$ tends to a Gaussian 
function~\cite{wan.87}. 
One can see from this example
that the function is never truly Gaussian - indeed it is always of the form
$\ln P \sim (3N/2 - 1) \ln E -\beta E$, which can easily be made independent 
of $N$ by choosing appropriate units. 
The central limit theorem applies because the width of the distribution 
scales as $N^{1/2}$ which means that fluctuations with any physical 
significance are all concentrated near the turning point of the function 
$\ln P$. 
The theorem only has meaning because of the significance one attaches to 
values of the variate that differ by only a few standard deviations from 
the mean. 
In practical terms it is therefore essential to normalise fluctuations to 
the standard deviation in order to test the central limit theorem.

In the case of dependent variables, the limit distribution can be different
from the Gaussian form. Two types of dependent random variable can be
defined \cite{cas.78}:
(i) weakly dependent, in which the correlation function falls to a constant
value in a finite range, and the standard deviation again varies as $\sqrt N$;
(ii) strongly dependent, in which the fluctuations vary as a power of $N$ 
different from one half.
Case (i) corresponds to a system with a finite correlation length.  
In case (ii), which includes systems with critical fluctuations, the 
central limit theorem does not hold, but a reasonable PDF can be obtained 
by normalising to the variance, hence to an appropriate power of $N$, with
$\rho \neq 0$.
Defining the (scalar) order parameter to be the intensive quantity $z=Z/N$, 
and using the scaling relations for a finite system, one finds 
$\rho = (1 - \eta/2)/D$. 
The limit distribution is now expected to be non-Gaussian, as can be shown
explicitly for the Ising model~\cite{gar.95,ell.85}.
Note however that, $\rho$ remains non-zero even at the upper critical 
dimension (taken as $D=4$ here), when $\eta =0$ and where one might 
legitimately expect a Gaussian PDF. 
The condition $\rho \ne 0$ may therefore be a necessary but not a sufficient 
condition to ensure non-Gaussian order parameter fluctuations.

Case (ii) is not actually limited to critical fluctuations: the
example of a dangerously irrelevant variable discussed in the text
also falls into this category, with $\rho = 2/D - 1/2$. 
Here, $\rho$ does go to zero as the upper critical dimension is reached and 
the danger of the irrelevant temperature variable disappears.

An ordinary critical point is more complicated than those of the
$2D$-XY model. In this case the correlation length is only infinite
precisely at the critical temperature. A non-Gaussian limit function
can therefore only be found on a locus of points such that $\xi/L$ is
a constant as the thermodynamic limit is taken. Thus, fixing
the temperature $T \ne T_C$ and varying $N$ will always cause a transition
from non-Gaussian to Gaussian statistics.  Conversely, fixing $T = T_C$
one will only arrive at the stable limit function in the thermodynamic limit.
One can therefore imagine a set of loci of constant PDF in $[T,L^{-1}]$ space 
that converge on $[T_C,0]$. 
We have suggested \cite{bra.00} that there is one such locus,
$[T^{\ast}(L),L^{-1}]$, where the PDF has approximately the same form as
that of the $2D$-XY model.
Thus, to sit at the critical temperature and change $L$ is
not the same as traveling along the
locus $[T^{\ast}(L),L^{-1}]$.
From scaling argument~\cite{bou.90} one can check that the tails of the PDF 
at $T_C$ should have the form $P(m)\sim \exp(-m^{\delta+1})$ in order to 
yield the correct scaling relation in the presence of a weak magnetic field:
$ \langle m \rangle \sim h^{1/\delta}$.
We do not find this, despite the same scaling relation holding for the
$2D$-XY model with $\delta = 8\pi J/k_{\rm B} T - 1$.
This difference may come from the difference in trajectories in the space of
variable $T$ and $L$.

A final point concerns the central limit theorem as applied to a vector order
parameter, ${\bf m}$, such as the XY model. 
In the high temperature limit, the fluctuations in the vector 
${\bf m}$ follow a two-dimension Gaussian centered on ${\bf m}=0$ and the PDF 
for the scalar $m = |{\bf m}|$ follows a ``Maxwell speed distribution'' for
a two-dimensional gas. 
In an ordered regime and even in the critical regime for $D=2$~\cite{arc.97}, 
$\sigma \ll \langle m \rangle $ which means $m$ behaves, to an excellent 
approximation, as a one-dimensional quantity. 
The symmetry breaking therefore induces a change in topology for the 
fluctuations in ${\bf m}$. 
This is generalisable to order parameters of higher dimension.

%%%%%%%%%%%%%%%%%%%%%%%%%%%%%%%%%%%%%%%
\section{}\label{app.1}
%%%%%%%%%%%%%%%%%%%%%%%%%%%%%%%%%%%%%%%

The graphs $g_k$ can be written,  in the large $N$ limit
in terms of power series. For example:
\be\label{eq.app1.1}
g_2=\lim_{N\rightarrow \infty}
\frac{4}{N^2}\sum_{m=1}^{Q}\sum_{n=1}^{Q}
\frac{1}{(4-2\cos 2\pi m/\sqrt{N}-2\cos 2\pi n/\sqrt{N})^2}
+\frac{4}{N^2}\sum_{m=1}^{Q}
\frac{1}{(4-2\cos 2\pi m/\sqrt{N})^2},
\ee
where $Q=(\sqrt{N}-1)/2$. The sum is dominated by the contributions
for small $m$ and $n$, but as the pole $m=0, \; n=0$ is explicitly excluded
from the sum, it remains finite even in the limit $N \rightarrow \infty$.
Taking only the first terms in a development of the cosines, which is exact in
the thermodynamic limit one finds
\be
g_2 &= &\frac{1}{4\pi^2}\sum_{m=1}^{\infty}
\frac{1}{m^4}
+\frac{1}{4\pi^4}\sum_{m=1}^{\infty}\sum_{n=1}^{\infty}
\frac{1}{(m^2+n^2)^2} \nonumber \\
&=&\frac{1}{360}+\frac{1}{4\pi^4}\sum_{m=1}^{\infty}\sum_{n=1}^{\infty}
\frac{1}{(m^2+n^2)^2},
\ee
and in general, for $g_k$
\be
g_k = \frac{1}{4\pi^2}\sum_{m=1}^{\infty}
\frac{1}{m^{2k}}
+\frac{1}{4\pi^4}\sum_{m=1}^{\infty}\sum_{n=1}^{\infty}
\frac{1}{(m^2+n^2)^k}.
\ee

%%%%%%%%%%%%%%%%%%%%%%%%%%%%%%%%%%%
\section{}\label{app.2}
%%%%%%%%%%%%%%%%%%%%%%%%%%%%%%%%%%%

For large and positive $y$ the functional form of $\varphi$ is 
$\varphi(y) \sim \frac{1}{8\pi}\log y + constant$.
To evaluate it in detail we use
the results of the App.~\ref{app.1} to write:
\be\label{eq.app2.1}
\varphi(y)=
\lim_{Q\rightarrow \infty}
\frac{1}{2\pi^2}\sum_{m=1}^{Q}\left (
\frac{1}{m^2}-\frac{1}{m^2+\hat y} \right )
+
\frac{1}{2\pi^2}\sum_{m=1}^{Q}\sum_{n=1}^{Q}\left (
\frac{1}{m^2+n^2}-\frac{1}{m^2+n^2+\hat y} \right ),
\ee
where $\hat y=y/4\pi^2$
The first two summations give, in the limit of large
$Q$, a constant and a function of $\hat y$ which tends to zero
for large argument:
\be
\lim_{y\rightarrow\infty}\frac{1}{2\pi^2}\sum_{m=1}^{\infty}\left (
\frac{1}{m^2}-\frac{1}{m^2+\hat y} \right )
=\frac{1}{12}
\ee
The double sum can be rewritten as
\be\label{eq.app2.2}
\frac{1}{2\pi^2}\sum_{m=1}^{Q}\sum_{n=1}^{Q}\left (
\frac{1}{m^2+n^2}-\frac{1}{m^2+n^2+\hat y} \right )
=
\frac{1}{2\pi^2}\sum_{m=1}^{Q}\sum_{n=1}^{\infty}\left (
\frac{1}{m^2+n^2}-\frac{1}{m^2+n^2+\hat y} \right )
-R(Q,y)
\ee
where $R$ is a correction term which vanishes in the limit of large $Q$:
\be
R(Q,y)=\frac{1}{2\pi^2}\sum_{m=1}^{Q}\sum_{n=Q+1}^{\infty}\left(
\frac{1}{m^2+n^2}-\frac{1}{m^2+n^2+\hat y} \right)
\ee
The sum can be evaluated in the limit $Q\rightarrow\infty$
using the Abel-Plana formula~\cite{olv.97}:
\be\label{eq.app2.3}
\sum_{i=p}^{q}f(i)=\int_p^qf(x)dx+\frac{1}{2}f(p)+\frac{1}{2}f(q)
+2\int_0^{\infty}\frac{\Im [f(q+ix)-f(p+ix)]}{\exp(2\pi x)-1}dx,
\ee
where $f$ is any real function that satisfied the assumptions 
in~\cite{olv.97}. 
Applying this to $R(Q,y)$, we have
\be \nn
R(Q-1,y)&=&
\frac{1}{2\pi^2} \sum_{m=1}^{Q-1}\frac{1}{m}\left[ \pi/2-\arctan(Q/m) \right]
+ \frac{1}{2(m^2+Q^2)}
+ 4Q\int_0^{\infty}\frac{xdx}{(x^2-m^2-Q^2)^2+4Q^2x^2}\frac{1}{\exp(2\pi x)-1}
\\ \nn
&-&\frac{1}{2\pi^2}
\sum_{m=1}^{Q-1}\frac{1}{\sqrt{m^2+\hat y}}
\left[ \pi/2-\arctan \left( Q/\sqrt{m^2+\hat y} \right) \right]
+\frac{1}{2(m^2+\hat y+Q^2)}
\\ \nn
&+&4Q\int_0^{\infty}\frac{xdx}{(x^2-m^2-\hat y-Q^2)^2+4Q^2x^2}\frac{1}{\exp
(2\pi x)-1}
\ee
The first term tends, in the large $Q$ limit, to the integral
\be
\sum_{m=1}^{Q-1}\frac{1}{m} 
\left[ \pi/2-\arctan(Q/m) \right] 
\rightarrow \int_0^{1}\frac{dx}{x} \left[ \pi/2-\arctan(1/x) \right]
=-\int_0^{1}\frac{\log x dx}{1+x^2}={\rm Catalan}.
\ee
A similar behaviour is found for the fourth term, since in this limit the 
dependence on $\hat y$ of this term vanishes as $\hat y/Q^2$. 
The other terms are corrections proportional to the inverse of some power 
of $Q$, so that $R$ vanishes in the large $Q$ limit.
The double sum (\ref{eq.app2.2}) can thus be reduced to a simple sum, since
\be\label{eq.app2.4}
\sum_{n=1}^{\infty}\frac{1}{n^2+z^2}=-\frac{1}{2z^2}+\frac{\pi}{2z}
\coth \pi z.
\ee
We therefore have, for large $y$
\be\nn
\frac{1}{2\pi^2}\sum_{m=1}^{Q}\sum_{n=1}^{\infty}\left (
\frac{1}{m^2+n^2}-\frac{1}{m^2+n^2+\hat y} \right )
&=&
\frac{1}{2\pi^2}\sum_{m=1}^{Q}-\frac{1}{2m^2}+\frac{\pi}{2m}
\coth \pi m
+\frac{1}{2(m^2+\hat y)}
\\ \nn
&-&\frac{\pi}{2\sqrt{m^2+\hat y}}
\coth \pi\sqrt{m^2+\hat y}
\\ \nn
&\simeq&
-\frac{1}{24}+\sum_{m=1}^{\infty}
\frac{1}{4\pi \sqrt{m^2+\hat y}} 
\left( 1-\coth \pi\sqrt{m^2+\hat y} \right)
+\frac{1}{4\pi m}(\coth \pi m-1)
\\ \label{eq.app2.5}
&+&\frac{1}{4\pi}\left (\frac{1}{m}-\frac{1}{\sqrt{m^2+\hat y}}\right ).
\ee
The series containing the hyperbolic function of $y$ vanishes in the
limit of large $y$ and the asymptotic behaviour of the last term can
be evaluated with the Abel-Plana formula (\ref{eq.app2.3})
\be\nn
\sum_{m=1}^{\infty}
\left (\frac{1}{m}-\frac{1}{\sqrt{m^2+\hat y}}\right )
&\simeq& \int_1^{\infty}dx\left (\frac{1}{x}-\frac{1}{\sqrt{x^2+\hat y}}
\right )
+\frac{1}{2}+2\int_0^{\infty}\frac{x\;dx}{(1+x^2)(\exp 2\pi x-1)}
\\ \label{eq.app2.6}
&=&\log \left (1+\sqrt{1+\hat y}\right )-\log 2 +\gamma
\ee
where the constant $\gamma$ is equal to
\be\nn
\gamma=\frac{1}{2}+2\int_0^{\infty}\frac{x\;dx}{(1+x^2)(\exp 2\pi x-1)}.
\ee
This can be proved by again applying the Abel-Plana formula to the
function $1/m$, since we know that $\sum_{m=1}^n1/m\simeq \log n +\gamma$.
The constant $\sum_m(1-\coth \pi m)/4\pi m$
in (\ref{eq.app2.5}) can be rewritten as
\be\nn
\sum_{m=1}^{\infty}\frac{1}{4\pi m}(\coth \pi m-1)&=&
\sum_{m=1}^{\infty}\frac{1}{2\pi m}\sum_{n=1}^{\infty}\exp(-2\pi mn)
\\ \label{eq.app2.7}
&=&-\frac{1}{2\pi}\log \prod_{n=1}^{\infty}\left[1-\exp(-2\pi n)\right]
\ee
and finally, the results (\ref{eq.app2.5},\ref{eq.app2.6},\ref{eq.app2.7})
give the asymptotic behaviour of $\varphi$ for large $y$:
\be
\varphi (y)=\frac{1}{8\pi}\log y+\frac{1}{24}-\frac{1}{4\pi}\log 4\pi
+\frac{\gamma}{4\pi}-\frac{1}{2\pi}\log \prod_{n=1}^{\infty}
\left[1-\exp(-2\pi n)\right]+\frac{1}{2y}+\cdots
\ee
The last term comes from a further study of the Abel-Plana formula
which gives the other correction terms in the inverse power of $y$.
An identical analysis gives the finite size magnetisation
\be
\langle m \rangle =\exp \left (-\frac{\tau}{2}\tr G/N \right )
\ee
 where $\tr G/N$ can be expand as
\be
\frac{1}{N}\tr G&=&\frac{1}{4\pi}\log CN \nonumber \\
 C & = &\exp\left\{
\frac{\pi}{3} + 2\log \frac{\sqrt{2}}{\pi}+ 2\gamma
- 4
\log \prod_{n=1}^{\infty}
\left[1-\exp(-2\pi n)\right] \right\} = 1.8456
\ee

\newpage

%%%%%%%%%%%%%%%%%%%%%%%%%%%%%%%%%%%%%%%%%%%%%%%%%%%%%%%%%%%%%%%%%%%%%%%%%%%%


\begin{references}
\bibitem{C} The constant $C$ is often quoted in the literature 
  with a value $C=2$~\cite{jos.77,tob.79,arc.97}.
  The more precise calculation here (App.~\ref{app.2}) leads to the 
  value $C=1.8456$.
\bibitem{fuckup} We note that we have made a similar change of variables for 
some of the data shown in ref.~\cite{bra.00} in order to have the exponential 
tail systematically on the left hand side of the figure.
The SOC data for the forest fire, Bak-Tang-Wiesenfeld and Sneppen models and 
the correlated extremal data are treated this way.
\end{references}
\end{document}